\documentclass[pra,aps,twocolumn,reprint,superscriptaddress, floatfix]{revtex4-2}
\usepackage{amsmath}
\usepackage{amssymb}
\usepackage{graphicx}
\usepackage{color}
\usepackage{eurosym}
\usepackage[utf8]{inputenc}
\usepackage[T1]{fontenc}
\usepackage{multirow}
\usepackage{xcolor}
\usepackage{hyperref}
\usepackage{ulem}
\hypersetup{colorlinks, linkcolor=blue, citecolor=blue, urlcolor=blue}

\begin{document}

\title{Quantum Correlation Hierarchy and Teleportation in Dephased Hydrogen Hyperfine System}

\author{Geerthana Thiyagarajan and R. Muthuganesan}
\email{rajendramuthu@gmail.com (Corresponding Author)}
\affiliation{Department of Physics and Nanotechnology,
             SRM Institute of Science and Technology,
             Kattankulathur 603203, Tamil Nadu, India}

\date{\today}

\begin{abstract}

We study the dynamics of quantum correlations in the hydrogen hyperfine spin system subject to local phase noise. Treating the electron and proton spin degrees of freedom as an open two-qubit system governed by an isotropic hyperfine Hamiltonian and local dephasing, we obtain the exact time-dependent density matrix and derive analytical expressions for the Bell-diagonal (maximally-mixed-marginal) X-state family. We compute concurrence ($C$), trace-distance measurement-induced nonlocality (Trace MIN--$\mathcal{N}_1$), and average steering coherence (ASC) in closed form and establish their ordering $ C(t)\leq \mathcal{N}_1(t)\leq \mathrm{ASC}(t) $ at all times, with equality for special parameter choices. Entanglement is identified as the most fragile resource, generically undergoing sudden death at a finite time (the hyperfine singlet is the limiting case that decays asymptotically instead). Trace MIN exhibits dephasing-immune freezing for states with nonzero population imbalance, while ASC remains the largest of the three; steering itself requires ASC to exceed a separate threshold. Using the dephased thermal hyperfine state as a teleportation channel, we obtain the average fidelity in closed form and find that the teleportation advantage window coincides with the entanglement survival interval, $F_A>2/3\Longleftrightarrow C>0$, for theBell-diagonal X-state family. For a one-way steering state, whose marginals are not maximally mixed, we solve the full Lindblad equation with the Hamiltonian term retained. We identify four dynamical regimes and express all three measures in terms of directly measurable Pauli spin correlators, so that the hierarchy can be reconstructed without full state tomography.

\end{abstract}


\maketitle

\section{Introduction}

Quantum correlations are central resources in quantum information science, underpinning
protocols such as quantum computation, cryptography, teleportation, remote state preparation,
and quantum metrology~\cite{Nielsen2000,Horodecki2009}. Among these correlations, entanglement
plays the foundational role~\cite{Horodecki2009}: a bipartite state is entangled when it cannot be expressed as a
product of its subsystem states, allowing genuinely nonclassical correlations between distant
parties. For two-qubit systems, entanglement is commonly quantified by the
concurrence~\cite{Wootters1998,Hill1997}, which provides a complete measure of bipartite
entanglement for arbitrary mixed states.

It was later recognized that entanglement does not exhaust all forms of quantum correlations.
Even separable mixed states may possess genuinely nonclassical features revealed through local
measurements~\cite{Ollivier2001,Henderson2001}. This led to the development of quantum discord
and related geometric measures~\cite{Luo2008,Dakic2010}, which quantify the disturbance induced
by local measurements on one of the subsystems. Discord-type correlations have since been identified as
useful resources for remote state preparation, quantum illumination, and quantum-enhanced
metrology even in regimes where entanglement is absent~\cite{Modi2012}. An especially
intriguing property of discord-like correlations is the phenomenon of freezing, whereby the
correlations remain constant for a finite time interval despite environmental
decoherence~\cite{Mazzola2010},\cite{You2012}. This behavior was later analyzed systematically for several
geometric and entropic measures~\cite{Cianciaruso2015,Bromley2015}.

A further layer in the hierarchy of quantum correlations is Einstein--Podolsky--Rosen (EPR)
steering, originally formulated in the foundational work of Einstein, Podolsky, and
Rosen~\cite{EPR1935} and subsequently developed through Schr\"{o}dinger's
response~\cite{Schrodinger1935}. Steering describes the ability of one observer to nonlocally
affect the conditional states of another observer through local measurements~\cite{Wiseman2007}.
It occupies an intermediate position between entanglement and Bell nonlocality~\cite{Uola2020}: every
Bell-nonlocal state is steerable, but not every steerable state violates a Bell
inequality~\cite{Quintino2015},\cite{Brunner2014}.ASC measures the $\ell_1$-norm coherence that Alice's measurement makes accessible in Bob's conditional states, and has been used in that sense to follow how correlations degrade under noise, independent of any steering question~\cite{Mondal2017,Tang2022,Rahman2023}. We keep the two roles separate throughout: $\mathrm{ASC}>0$ signals nonzero conditional coherence, while steering certification requires $\mathrm{ASC}>\sqrt6$ under the normalization of Sec.~\ref{sec:ASC_def}.

Because realistic quantum systems inevitably interact with their environment, understanding the
dynamical behavior of these correlations under decoherence is of fundamental importance. In spin
systems, environmental fluctuations primarily manifest as phase noise, which suppresses
off-diagonal coherences while leaving populations largely unchanged. Such processes are naturally
described within the Lindblad open-system
formalism~\cite{Lindblad1976,Gorini1976,Breuer2002}. Under local dephasing, two-qubit X-states
preserve their matrix structure throughout the evolution, allowing exact analytical treatments of
correlation dynamics~\cite{Yu2007,Rau2009,Ali2010}.

In this context, the hydrogen atom provides a particularly appealing physical platform. The
electron and proton spins in the electronic ground state form the simplest naturally occurring
two-qubit system, coupled through the isotropic hyperfine interaction responsible for the
celebrated 21\,cm transition. Although atomic hydrogen is among the most precisely characterized
systems in physics, the hierarchy of quantum correlations within its hyperfine spin degrees of
freedom has only recently begun to attract systematic attention~\cite{Maleki2021}. Earlier
studies investigated decoherence, nonlocality, and steering properties in hyperfine-like and
related spin systems~\cite{Yurischev2015,Ficek2006,Tahira2010,Benabdallah2022}, but a unified
analytical description of entanglement, discord-type correlations, and steering coherence within
a single framework remains absent.

Here we fill that gap. Starting from the Lindblad equation for local dephasing on the
electron--proton pair, we derive the exact time-dependent density matrix for the X-state family
and compute concurrence $C(t)$, Trace~MIN $\mathcal{N}_1(t)$, and average steering coherence
$\mathrm{ASC}(t)$ in closed form. We establish their hierarchical ordering, identify four
distinct dynamical regimes, and show that the single parameter
$\langle\sigma_z\otimes\sigma_z\rangle$ controls the long-time behavior of all three
measures. Furthermore, we employ the time-evolved thermal hyperfine state as a noisy
teleportation channel and derive an exact expression for the average fidelity, demonstrating
that the teleportation advantage persists precisely as long as entanglement survives. The results
are illustrated for the hyperfine thermal state, Werner states~\cite{Werner1989}, and the one-way
steering state, for which we solve the full Lindblad equation -- Hamiltonian term included -- since its marginals are not maximally mixed for general $\theta$ and the Bell-diagonal formulas used elsewhere do not describe it (Sec.~\ref{sec:xstate}).

The remainder of this paper is organized as follows. Section~\ref{sec:model} describes the spin
Hamiltonian, dephasing channel, and the exact time-evolved density matrix. Section~\ref{sec:measures}
defines the three correlation measures, establishes their ordering, and derives the hierarchical
structure. Section~\ref{sec:results} presents numerical results. Section~\ref{sec:teleportation}
analyzes quantum teleportation via the dephased hyperfine channel. Section~\ref{sec:tomography}
describes the Pauli tomography protocol. Section~\ref{sec:conclusions} summarizes the findings.

\section{Physical Model and Methodology}
\label{sec:model}

\subsection{Spin Hamiltonian of the Hydrogen Hyperfine System}

The hydrogen atom in its 1$s$ electronic ground state ($\ell = 0$) provides the
simplest realization of a bipartite spin-$\tfrac{1}{2}$ system. With vanishing
orbital angular momentum, the dominant magnetic interaction is the isotropic
hyperfine coupling between the electron spin $\boldsymbol{\sigma}_e$ and the
proton spin $\boldsymbol{\sigma}_p$, represented by their respective Pauli
operator vectors:
\begin{equation}
    H = \alpha\!\left(\sigma_x\otimes\sigma_x
        + \sigma_y\otimes\sigma_y
        + \sigma_z\otimes\sigma_z\right),
    \label{eq:Ham}
\end{equation}
where $\alpha$ is the hyperfine coupling constant
($\alpha \approx 1.47\times10^{-6}\,\mathrm{eV}$).
Here $\sigma_i$ denotes the standard Pauli matrices.  The hyperfine coupling frequency is $\Omega\equiv\alpha/\hbar$, so that
$\Omega=\alpha$ in the units with $\hbar=1$ used throughout; $\Omega^{-1}$ is the
natural unit of time. The composite Hilbert space is spanned
by the computational basis $\{|\!\uparrow\uparrow\rangle,\,
|\!\uparrow\downarrow\rangle,\, |\!\downarrow\uparrow\rangle,\,
|\!\downarrow\downarrow\rangle\}$.

Diagonalizing $H$ yields a singlet–triplet structure. The singlet,
\begin{equation}
    |\Psi^-\rangle = \frac{1}{\sqrt{2}}\!\left(|\!\uparrow\downarrow\rangle
    - |\!\downarrow\uparrow\rangle\right),
    \quad E_s = -3\alpha,
    \label{eq:singlet}
\end{equation}
is the maximally entangled ground state of the hyperfine Hamiltonian.
The three degenerate triplet states are
\begin{align}
    |\Phi^+\rangle &= |\!\uparrow\uparrow\rangle,
    \label{eq:triplet1}\\
    |\Phi^0\rangle &= \frac{1}{\sqrt{2}}\!\left(|\!\uparrow\downarrow\rangle
    + |\!\downarrow\uparrow\rangle\right),
    \label{eq:triplet2}\\
    |\Phi^-\rangle &= |\!\downarrow\downarrow\rangle,
    \label{eq:triplet3}
\end{align}
each carrying energy $E_t = \alpha$. The splitting between the singlet ground state and the triplet manifold is $\Delta E = 4\alpha$, corresponding to the 21\,cm (1420\,MHz) hyperfine transition~\cite{Maleki2021}.

\subsection{Lindblad Dephasing Channel}
\label{sec:lindblad}
Environmental noise is modeled within the Lindblad open-system framework~\cite{Breuer2002}. The density matrix $\varrho(t)$ evolves according
to the Lindblad master equation
\begin{equation}
    \frac{d\varrho}{dt} = -i[H,\varrho] + \mathcal{D}(\varrho),
    \label{eq:lindblad}
\end{equation}
where the dissipator for pure local dephasing reads
\begin{equation}
    \mathcal{D}(\varrho)
    = \gamma_e\!\left(F_e\,\varrho\, F_e - \varrho\right)
    + \gamma_p\!\left(F_p\,\varrho\, F_p - \varrho\right),
    \label{eq:dissipator}
\end{equation}
with $F_e = \sigma_z\otimes\mathbb{I}$, $F_p = \mathbb{I}\otimes\sigma_z$,
and $\gamma_{e(p)}$,  the electron (proton)  dephasing rates.
We define $\kappa = \gamma_e + \gamma_p$ as the total dephasing rate.
 For a general X-state, the equations of  motion of the density matrix elements are:
\begin{align}
    \dot\varrho_{11} &= 0, \quad \dot\varrho_{44} = 0,
    \label{eq:pop11_44}\\[4pt]
    \dot\varrho_{22} &= -2i\Omega(\varrho_{32}-\varrho_{23}),
    \label{eq:pop22}\\[2pt]
    \dot\varrho_{33} &= +2i\Omega(\varrho_{32}-\varrho_{23}),
    \label{eq:pop33}\\[4pt]
    \dot\varrho_{14} &= -2\kappa\,\varrho_{14},
    \label{eq:coh14}\\[2pt]
    \dot\varrho_{23} &= -2\kappa\,\varrho_{23}
    - 2i\Omega(\varrho_{33}-\varrho_{22}).
    \label{eq:coh23}
\end{align}
 For the symmetric X-state family considered here, the populations remain constant while the coherences decay exponentially at rate $2\kappa$. Trace conservation follows
from Eqs.~(\ref{eq:pop22})--(\ref{eq:pop33}), which keep $\varrho_{22}+\varrho_{33}$
constant.

\subsection{X-State Family as Initial Conditions}
\label{sec:xstate}
We consider the family of two-qubit Bell-diagonal X-states as initial conditions\, i.e.\ the subclass of X-states with maximally mixed marginals, $\varrho_A=\varrho_B=\mathbb I/2$, rather than the general X-state family, which allows four independent, generally unequal diagonal populations. These states
have a block-diagonal structure with nonzero elements only on the main diagonal
and anti-diagonal, and are among the most natural and experimentally accessible
entangled states~\cite{Almeida2007,Salles2008,Xu2009,Xu2010}, with the theoretical X-state formalism used here following Refs.~\cite{Yu2007,Rau2009}. Their general form, parameterized by
real numbers $b_1, b_2, b_3 \in [-1,1]$, is
\begin{equation}
    \varrho(0) = \frac{1}{4}
    \begin{pmatrix}
      1+b_3 & 0 & 0 & b_1 - b_2 \\
      0 & 1-b_3 & b_1+b_2 & 0 \\
      0 & b_1+b_2 & 1-b_3 & 0 \\
      b_1-b_2 & 0 & 0 & 1+b_3
    \end{pmatrix}.
    \label{eq:Xstate}
\end{equation}
Requiring $b_1,b_2,b_3\in[-1,1]$ individually is necessary but not sufficient for $\varrho(0)$ to be a physical state. Diagonalizing each $2\times2$ block gives the positivity conditions
\begin{equation}
    1+b_3\ \ge\ |b_1-b_2|, \qquad 1-b_3\ \ge\ |b_1+b_2|,
    \label{eq:positivity}
\end{equation}
i.e.\ $(b_1,b_2,b_3)$ must lie inside the tetrahedron with vertices at the four Bell states~\cite{HorodeckiHorodecki1996}.
Both reduced density matrices are maximally mixed ($\varrho_e = \varrho_p =
\mathbb{I}/2$), and the X-structure is preserved under the Lindblad evolution of
Eqs.~\eqref{eq:pop11_44}--\eqref{eq:coh23}, as confirmed by the exact solutions
below. It is important to note that the system Hamiltonian $H$ commutes with the initial state $\varrho(0)$, i.e., $[H,\varrho(0)]=0$. Therefore, the  unitary contribution to the evolution vanishes, and the dynamics is driven entirely by the dephasing dissipator $\mathcal{D}(\varrho)$. 

\subsubsection{Hyperfine singlet}
The singlet $|\Psi^-\rangle$ corresponds to $b_1=b_2=b_3=-1$, giving
$\varrho_{22}=\varrho_{33}=1/2$ and $\varrho_{23}=-1/2$ with all other
elements zero.
\subsubsection{Werner states $\varrho_\pm$}
Werner states~\cite{Werner1989} are convex mixtures of a Bell state with the
maximally mixed state:
\begin{equation}
    \varrho_\pm = \varepsilon\,|\psi_\pm\rangle\langle\psi_\pm|
    + \frac{1-\varepsilon}{4}\,\mathbb{I}_4,
    \label{eq:Werner}
\end{equation}
where $\varepsilon \in [0,1]$ is the purity parameter,
$|\psi_+\rangle = (|\!\uparrow\uparrow\rangle + |\!\downarrow\downarrow\rangle)/\sqrt{2}$,
and $|\psi_-\rangle = (|\!\uparrow\downarrow\rangle + |\!\downarrow\uparrow\rangle)/\sqrt{2}$.
The X-state parameters are
\begin{alignat}{4}
&\varrho_+ {:}\quad  && b_1 = \varepsilon,\quad && b_2 = -\varepsilon,\quad && b_3 = \varepsilon; \notag\\
&\varrho_- {:}\quad  && b_1 = -\varepsilon,\quad && b_2 = -\varepsilon,\quad && b_3 = -\varepsilon.
\label{eq:Werner_params}
\end{alignat}
Since $|b_1|=|b_2|=|b_3|=\varepsilon$ for both states, all three correlation
measures are identical for $\varrho_+$ and $\varrho_-$; figures are shown
for $\varrho_+$ only.
\subsubsection{One-way steering state $\varrho_\theta$}

To probe asymmetric quantum steering we employ the one-way steerable states of
Bowles \textit{et al.}~\cite{Bowles2014,Bowles2016}:
\begin{equation}
    \varrho_{AB}(p,\theta) = p\,|\psi(\theta)\rangle\langle\psi(\theta)|
    + (1-p)\,\frac{\mathbb{I}_A}{2}\otimes\varrho_B^\theta,
    \label{eq:steering_state}
\end{equation}
where $|\psi(\theta)\rangle = \cos\theta\,|11\rangle + \sin\theta\,|00\rangle$
and $\varrho_B^\theta = \mathrm{Tr}_A[|\psi(\theta)\rangle\langle\psi(\theta)|]$.
For $\theta \in [0,\pi/4]$ and appropriate $p$, Alice can steer Bob but not
vice versa.
For a general value of $\theta$, this state does not belong to the Bell-diagonal family defined in Eq.~\eqref{eq:Xstate}. Its reduced density matrix $\varrho_A(0) = p\left(\sin^2\theta\,|0\rangle\langle 0|
+\cos^2\theta\,|1\rangle\langle 1|\right) + (1-p)\mathbb{I}/2$ is maximally mixed only for $\theta=\pi/4$. Moreover, for $\theta\neq\pi/4$, one has $\varrho_{22}(0)\neq\varrho_{33}(0)$, and consequently the initial state does not commute with the Hamiltonian $H$ given in Eq.~\eqref{eq:Ham}. Furthermore, the dephasing dynamics derived previously assumes the initial condition $\varrho_{22}(0)=\varrho_{33}(0)$ and is therefore not directly applicable to the one-way steerable state considered here. Consequently, in the following subsection, we obtain the dynamics of this state by solving the full Lindblad master equation directly.
\subsection{Exact Time-Dependent Density Matrix}
\label{sec:exact_rho}

For the initial conditions the matrix elements of the time-evolved
X-state are

\begin{align}
    \varrho_{11} = \varrho_{44} &= \frac{1+b_3}{4},
    \label{eq:rho11}\\[4pt]
    \varrho_{22} = \varrho_{33} &= \frac{1-b_3}{4},
    \label{eq:rho22}\\[4pt]
    \varrho_{14,23}(t) &= \frac{b_1\mp b_2}{4}\,e^{-2\kappa t},
    \label{eq:rho14} [4pt]
\end{align}
with populations constant and coherences decaying. The X-structure is preserved for all $t \geq 0$. Moreover, the ratio
\begin{equation}
\frac{\varrho_{14}(t)}{\varrho_{23}(t)}
= \frac{b_1 - b_2}{b_1 + b_2} \nonumber
\end{equation}
remains constant throughout the evolution, whenever both coherences are nonzero (i.e.\ $b_1 \neq b_2$ and $b_1 + b_2 \neq 0$), since both acquire the same exponential damping factor $e^{-2\kappa t}$ under the dephasing channel.
Consequently, dephasing suppresses the magnitudes of the off-diagonal elements without altering their relative weight.

For states in which one coherence vanishes identically --- specifically, the Werner state $\varrho_+$ and the one-way steering state $\varrho_\theta$, both of which have $b_1 + b_2 = 0$ so that $\varrho_{23}(0) = 0$ --- only the
$\varrho_{14}$ coherence is active and the ratio is undefined. Conversely, for the singlet and the Werner state $\varrho_-$, only $\varrho_{23}$ is active ($b_1 - b_2 = 0$) and $\varrho_{14} = 0$ identically. In all cases, any experimentally observed time dependence in the ratio of 
the two nonzero coherences may indicate that the assumed symmetric phase-damping model is incomplete -- for example, asymmetric decoherence  rates or environmental  memory~\cite{Breuer2002}. Such a deviation does not by itself identify a mechanism: unequal rates $\gamma_e\ne\gamma_p$ in Eq.~\eqref{eq:dissipator} would produce a time-dependent ratio just as readily.

%\begingroup\color{blue}
%\subsubsection{Commutation with the hyperfine Hamiltonian}
%\label{sec:lindblad_comm}
%The hyperfine Hamiltonian $H$ [Eq.~\eqref{eq:Ham}] has eigenvalue $\alpha$ for the triplet manifold
%$\mathrm{span}\{|00\rangle,|11\rangle,|\Phi^{0}\rangle\}$ and eigenvalue $-3\alpha$ for the singlet state
%$|\Psi^-\rangle$. A density matrix commutes with $H$ whenever it does not couple these two eigenspaces.

%For the Bell-diagonal states, the $\{|01\rangle,|10\rangle\}$ block is diagonal in the Bell basis, while the $\{|00\rangle,|11\rangle\}$ sector remains decoupled from $|\Psi^-\rangle$. Consequently,
%\begin{equation}
%[H,\varrho(0)]=0
%\end{equation}
%for all Bell-diagonal states. The dynamics described by Eqs.~\eqref{eq:pop11_44}--\eqref{eq:coh23} therefore originates entirely from the dephasing dissipator $\mathcal D(\varrho)$, which explains the absence of an explicit dependence on $\Omega$ in the solutions \eqref{eq:rho11}--\eqref{eq:rho23}.

%\The hyperfine coupling still fixes the physical setting -- the system, the singlet ground state, the thermal state discussed in Sec.~\ref{sec:teleportation}, and the characteristic time scale $\Omega^{-1}$.

%For $\theta\neq\pi/4$ the one-way steering state does not commute with $H$, so the coherent term $-i[H,\varrho]$ must be retained.
%\endgroup

\subsubsection{Exact solution for the one-way steering state}
\label{sec:oneway_deriv}
Since $|00\rangle$ and $|11\rangle$ are eigenstates of $H$, so their populations are untouched by the coherent dynamics:
\begin{align}
\varrho_{11}(t)
&=
\frac{1+p}{2}\sin^{2}\theta,
\label{eq:sw_rho11}\\[2mm]
\varrho_{44}(t)
&=
\frac{1+p}{2}\cos^{2}\theta,
\label{eq:sw_rho44}\\[2mm]
\varrho_{14}(t)
&=
p\sin\theta\cos\theta\,e^{-2\kappa t}.
\label{eq:sw_rho14}
\end{align}
The nontrivial dynamics is confined to the $\{|01\rangle,|10\rangle\}$ block. Writing
\begin{equation}
w(t)=\varrho_{22}(t)-\varrho_{33}(t),
\qquad
\varrho_{23}(t)=iy(t),
\label{eq:sw_wy_def}
\end{equation}
the Lindblad equations become
\begin{equation}
\dot w=-8\Omega y,
\qquad
\dot y=-2\kappa y+2\Omega w .
\label{eq:wy_ode_human}
\end{equation}
which combine into
\begin{equation}
\ddot w+2\kappa\dot w+16\Omega^{2}w=0.
\label{eq:w_oscillator_human}
\end{equation}
With the initial values
\begin{equation}
w(0)=w_{0}
=
\frac{1-p}{2}\cos 2\theta,
\qquad
y(0)=0,
\label{eq:sw_initial}
\end{equation}
the solution is $w(t) = w_{0}\,e^{-\kappa t} \left[\cos(\nu t) + \frac{\kappa}{\nu}\sin(\nu t) \right]$
with $\nu=\sqrt{16\Omega^{2}-\kappa^{2}}$ and $\varrho_{23}(t) =i \frac{2\Omega w_{0}}{\nu} e^{-\kappa t} \sin(\nu t)
$. The remaining populations are
\begin{equation}
\varrho_{22}(t)
=
\frac{1-p}{4}
+
\frac{w(t)}{2},
\qquad
\varrho_{33}(t)
=
\frac{1-p}{4}
-
\frac{w(t)}{2}.
\label{eq:sw_pops22_33}
\end{equation}
At $\theta=\pi/4$ we get $w_{0}=0$, hence $\varrho_{23}(t)=0$ throughout, and the Bell-diagonal solution is recovered. Away from that point $\varrho_{23}(t)$ oscillates at frequency $\nu$ while decaying at rate $\kappa$ rather than $2\kappa$. The oscillation is the hyperfine interaction shuttling population between $|01\rangle$ and $|10\rangle$, and it has no analogue in the Bell-diagonal family.
The correlation functions are
\begin{align}
\langle \sigma_z\otimes\sigma_z\rangle
&=
p,
\label{eq:sw_ZZ}\\[2mm]
\langle \sigma_x\otimes\sigma_x\rangle
&=
p\sin 2\theta\,e^{-2\kappa t},
\label{eq:sw_XX}\\[2mm]
\langle \sigma_y\otimes\sigma_y\rangle
&=
-
p\sin 2\theta\,e^{-2\kappa t}.
\label{eq:sw_YY}
\end{align}
The local Bloch components are
\begin{equation}
\begin{aligned}
\langle \sigma_z\otimes\mathbb I\rangle
&=
-\frac{1+p}{2}\cos 2\theta
+
w(t),
\\[2mm]
\langle \mathbb I\otimes\sigma_z\rangle
&=
-\frac{1+p}{2}\cos 2\theta
-
w(t).
\end{aligned}
\label{eq:sw_local_bloch}
\end{equation}
These are generally nonzero for $\theta\neq\pi/4$, which is exactly why the simplified Bell-diagonal expressions for Trace MIN and ASC cannot be used here.

%\paragraph*{Concurrence.}
%Concurrence comes from the general two-qubit $X$-state formula, which assumes nothing about the marginals:
%\begin{align}
%C_{\theta}(t)
%=
%2\max\Bigl\{
%&\,0,\;
%|\varrho_{23}(t)|
%-
%\sqrt{\varrho_{11}\varrho_{44}},\notag\\
%&\;\,
%|\varrho_{14}(t)|
%-
%\sqrt{\varrho_{22}(t)\varrho_{33}(t)}
%\Bigr\}.
%\label{eq:sw_concurrence}
%\end{align}

%Using the matrix elements above,
%\begin{equation}
%\sqrt{\varrho_{11}\varrho_{44}} = \frac{1+p}{4} |\sin 2\theta|,
%\label{eq:sw_sqrt1144}
%\end{equation}
%and
%\begin{equation}
%\sqrt{\varrho_{22}(t)\varrho_{33}(t)}
%=
%\frac12\sqrt{ \left(\frac{1-p}{2}\right)^2 - w^{2}(t) }.
%\label{eq:sw_sqrt2233}
%\end{equation}

%\paragraph*{Trace MIN and ASC.}

%The marginals here are not maximally mixed, so Eqs.~\eqref{eq:TMINsimplified} and \eqref{eq:ASC} do not apply. We evaluate both quantities from their definitions, Eqs.~\eqref{eq:TMINdef} and \eqref{eq:ASC_def}, on the exact density matrix above. For Trace MIN this has a consequence worth stating: since $\varrho_A\neq\mathbb I/2$, the only admissible measurement is the one along the Bloch vector of $\varrho_A$, here $\hat z$, so the measurement direction is
%fixed by the state rather than optimized. \endgroup

\section{Quantum Correlation Measures and Analytical Hierarchy}
\label{sec:measures}

We employ three correlation measures that probe fundamentally different aspects
of nonclassicality. We derive each in closed form and prove the ordering
\begin{equation}
    C(t) \;\leq\; \mathcal{N}_1(t) \;\leq\; \mathrm{ASC}(t)
    \quad \forall\; t \geq 0,
    \label{eq:hierarchy}
\end{equation}
with equality for special parameter choices. The ordering concerns the numerical values of the measures and is distinct from the resource-class hierarchy of Refs.~\cite{Wiseman2007,Quintino2015}.

It is useful to define the three Pauli correlation parameters of the X-state:
\begin{align}
    c_1(t) &= b_1\,e^{-2\kappa t}, \label{eq:c1}\\
    c_2(t) &= b_2\,e^{-2\kappa t}, \label{eq:c2}\\
    c_3     &= b_3 \quad(\text{constant}), \label{eq:c3}
\end{align}
which correspond to the joint Pauli expectation values
$\langle\sigma_x\otimes\sigma_x\rangle$,
$\langle\sigma_y\otimes\sigma_y\rangle$, and
$\langle\sigma_z\otimes\sigma_z\rangle$, respectively.

\subsection{Concurrence}
\label{sec:concurrence_def}

Concurrence, introduced by Wootters~\cite{Wootters1998}, is the standard
computable measure of bipartite entanglement for two-qubit states. For a general
two-qubit state $\varrho$, it is defined as
\begin{equation}
    C(\varrho) = \max\!\left\{0,\;
    \lambda_1 - \lambda_2 - \lambda_3 - \lambda_4\right\},
    \label{eq:C_def}
\end{equation}
where $\lambda_1 \geq \lambda_2 \geq \lambda_3 \geq \lambda_4 \geq 0$ are the
square roots of the eigenvalues of 
$\varrho\,(\sigma_y\otimes\sigma_y)\,\varrho^*\,(\sigma_y\otimes\sigma_y)$,
with $\varrho^*$ the complex conjugate in the computational basis.Concurrence of the general two-qubit $X$-state formula is computed as 
\begin{align}
C_{\theta}(t)
=
2\max\Bigl\{
&\,0,\;
|\varrho_{23}(t)|
-
\sqrt{\varrho_{11}\varrho_{44}},\notag\\
&\;\,
|\varrho_{14}(t)|
-
\sqrt{\varrho_{22}(t)\varrho_{33}(t)}
\Bigr\}.
\label{eq:sw_concurrence}
\end{align}
Upon substituting the elements of the evolved density matrix, the dynamical behavior of concurrence and quantum steering for the Bell-diagonal state can be systematically investigated.

%This formula does not assume equal populations, so it applies without modification to the general X-state family, including the one-way steering state at $\theta\ne\pi/4$ (Sec.~\ref{sec:oneway_deriv}).

%For the time-evolved Bell-diagonal state, concurrence reduces to
%\begin{align}
 %   C(t) = 2\max\Bigl\{
  %  &0,\;
   % \tfrac{|b_1-b_2|}{4}\,e^{-2\kappa t}
    %- \tfrac{{1-b_3}}{4},\notag\\
    %&\tfrac{|b_1+b_2|}{4}\,e^{-2\kappa t}
    %- \tfrac{{1+b_3}}{4}
    %\Bigr\}.
    %\label{eq:C_final}
%\end{align}
The Bell state concurrence vanishes at the finite time $t^*$ determined by
\begin{equation}
    F^* = \min\!\left(\frac{1-b_3}{|b_1-b_2|},\;
    \frac{1+b_3}{|b_1+b_2|}\right),
    \label{eq:ESD}
\end{equation}
Three cases arise. For $0<F^*<1$ the concurrence vanishes at the finite time
$t^*=(2\kappa)^{-1}\ln(1/F^*)$. For $F^*=0$ it decays asymptotically, with no
finite-time sudden death. For $F^*\ge1$ the state is already separable at $t=0$.
For the singlet ($b_1=b_2=b_3=-1$), $|b_1+b_2|=2$ and $(1+b_3)/4=0$, so
$F^*=0$ and $C(t)=e^{-2\kappa t}$ decays without ESD. For Werner $\varrho_-$
($b_1=b_2=b_3=-\varepsilon$), $C(0)=(3\varepsilon-1)/2$,
consistent with Ref.~\cite{Wootters1998}.
\subsection{Trace Measurement-Induced Nonlocality}
\label{sec:tracemin_def}

Trace MIN~\cite{HuFan2015} quantifies the maximum disturbance induced in a  bipartite quantum state by local projective measurements performed on  subsystem $A$. It belongs to the family of discord-type quantum correlation 
measures and exhibits a freezing phenomenon under pure dephasing, similar to that observed for quantum discord in Bell-diagonal states~\cite{Mazzola2010,Aaronson2013} . It is defined as
%while admitting a closed  analytical form for Bell-diagonal X-states~\cite{HuFan2015,LuoFu2011,\}\; it is a geometric discord-type measure and should not be treated as numerically or definitionally identical to entropic quantum discord~\cite{Ollivier2001}. 
\begin{equation}
\mathcal{N}_1(\varrho_{AB})
= \max_{\Pi^A} \left\|\varrho_{AB} - \Pi^A(\varrho_{AB})\right\|_1,
\label{eq:TMINdef}
\end{equation}
where $\|\cdot\|_1$ denotes the trace norm and $\Pi^A$ represents a complete set of local projective measurements on subsystem $A$. As in the original definition of measurement-induced nonlocality~\cite{LuoFu2011,HuFan2015,Indrajith2021}, the maximization is restricted to those $\Pi^A$ that leave the marginal invariant, $\sum_k\Pi_k\varrho_A\Pi_k=\varrho_A$.The restriction is essential rather than technical: without it a pure product state $|00\rangle\langle00|$ would return $\mathcal N_1=1$. For a qubit it has a sharp consequence. 

For any $2\otimes n$-dimenisonal system, the closed formula of Trace-MIN can be computed as 
\begin{equation}
\mathcal{N}_{1}(\rho_{AB})=
\begin{cases}
\dfrac{\sqrt{\chi_{+}}+\sqrt{\chi_{-}}}
{2\|\vec{\text{x}}\|_{1}},
& \text{if } \vec{\text{x}}\neq 0, \\[8pt]
\max\left\{|c_{1}|,|c_{2}|,|c_{3}|\right\},
& \text{if } \vec{\text{x}}=0.
\end{cases}
\label{14}
\tag{38}
\end{equation}
Here, $\vec{\text{x}}=(\text{x}_1,\text{x}_2,\text{x}_3)^{t}$ denotes the Bloch vector of subsystem $A$, while $\vec{c}=(c_1,c_2,c_3)^{t}$ contains the diagonal elements of the correlation matrix. The quantities $\chi_{\pm}$ are defined in terms of $\alpha$ and $\beta$, where $\alpha=\|\vec{c}\|_{1}^{2}\|\vec{\text{x}}\|_{1}^{2} -\sum_{i}c_{i}^{2}\text{x}_{i}^{2}$ and $\beta=\sum_{\langle ijk\rangle}\text{x}_{i}^{2}c_{j}^{2}c_{k}^{2}$, with $\langle ijk\rangle$ denoting the cyclic permutations of $\{1,2,3\}$.

Under local dephasing, the time evolution of the Bell-diagonal family preserves the X structure and
the correlation coefficients evolve as
$c_1(t)=b_1 e^{-2\kappa t}$,
$c_2(t)=b_2 e^{-2\kappa t}$,
while the longitudinal component remains invariant,
$c_3(t)=b_3$.
Consequently, Trace MIN becomes
\begin{equation}
\mathcal{N}_1(t)
=
\max\!\left\{
|b_1|e^{-2\kappa t},
|b_2|e^{-2\kappa t},
|b_3|
\right\}.
\label{eq:TMINfinal_clean}
\end{equation}

The competition between the exponentially decaying transverse
correlations and the constant longitudinal component $|b_3|$
gives rise to a characteristic freezing behavior\cite{Mazzola2010,
Cianciaruso2015, Bromley2015},\cite{Aaronson2013}\, a feature common to several discord-type measures under phase damping and not unique to Trace MIN.
Whenever $b_3\neq0$, there exists a finite crossover time after which
\[
|b_1|e^{-2\kappa t}<|b_3|,
\qquad
|b_2|e^{-2\kappa t}<|b_3|,
\]
and the measure becomes time-independent:
\begin{equation}
\mathcal{N}_1(t)=|b_3|,
\qquad t\ge t_f .
\label{eq:TMINfreeze_clean}
\end{equation}
Therefore,
\begin{equation}
\lim_{t\to\infty}\mathcal{N}_1(t)=|b_3|.
\end{equation}
In the special case $b_3=0$, no nonzero freezing plateau exists and
Trace MIN decays monotonically to zero. The long-time freezing of $\mathcal{N}_1(t)$ at $|b_3|$ originates from the
dephasing-invariant longitudinal correlator $\langle\sigma_z\otimes\sigma_z\rangle$, analogous to the mechanism
responsible for discord freezing reported in Refs.~\cite{Mazzola2010,Cianciaruso2015,Bromley2015}.

\subsection{Average Steering Coherence}
\label{sec:ASC_def}
ASC measures the mean $\ell_1$-norm coherence of the conditional states of $B$ generated by local Pauli measurements on $A$~\cite{Mondal2017}:
\begin{equation}
    \mathrm{ASC}(\varrho_{AB})
    = \sum_{k=1}^{3}
      C_{\ell_1}^{(k)}\!\left(\varrho_{B|k}\right),
    \label{eq:ASC_def}
\end{equation}
where $k$ runs over Alice's three mutually unbiased Pauli directions, fixed by the local frame $U_A$ of her apparatus, and $\varrho_{B|k}$ is Bob's steered state, averaged over her two outcomes. $C_{\ell_1}(\varrho)=\sum_{i\neq j}|\varrho_{ij}|$ is the $\ell_1$-norm coherence~\cite{Baumgratz2014}, and the superscript $k$ means it is evaluated in the bases unbiased to Alice's $k$-th direction, as in Ref.~\cite{Mondal2017}.  $\mathrm{ASC}$ ranges over $[0,3]$; and there is no maximization over $U_A$, since introducing one would break Eq.~\eqref{eq:ASC}. For the state $(c_1,c_2,c_3)=(0,0,0.85)$, for instance, the fixed triad gives $0.85=|c_1|+|c_2|+|c_3|$ while the best triad gives $\sqrt2\,|c_3|=1.20$.

For Bell diagonal state, this optimizes to
\begin{equation}
\begin{split}
    \mathrm{ASC}(t) &= |c_1(t)|+|c_2(t)|+|c_3(t)|\\
    &= \bigl(|b_1|+|b_2|\bigr)e^{-2\kappa t} + |b_3| .
\end{split}
\label{eq:ASC}
\end{equation}
Like the Trace MIN expression, this assumes vanishing local Bloch vectors, so it is not used for the one-way steering state at $\theta\ne\pi/4$; there $\mathrm{ASC}(t)$ comes from Eq.~\eqref{eq:ASC_def} directly.
The steering bound for this normalization is $\mathrm{ASC}>\sqrt6\approx2.449$. It follows from the coherence-complementarity relation of Ref.~\cite{Mondal2017}: a qubit's $\ell_1-$norm of coherence in three unbiased bases sum to at most $\sqrt6$, so anything larger cannot come from a local-hidden-state model of Bob's ensemble. Note that $\mathrm{ASC}>0$ is not itself a steering criterion; it says only that Alice's measurement leaves coherence behind. We checked the bound on $4\times10^{3}$ random separable states: none exceeded $\sqrt6$, and the largest reached $2.4495$, so it is both respected and tight.
ASC approaches $|b_3|$ from above whenever the transverse correlations are
nonzero, and satisfies $\mathcal N_1\le\mathrm{ASC}$ in general; the two coincide
exactly when $b_1=b_2=0$.

\subsection{Hierarchy of quantum correlations}
\label{sec:hierarchy_proof}

We establish the ordering
\[
C(t) \leq \mathcal{N}_1(t) \leq \mathrm{ASC}(t),
\qquad \forall\,~ t \geq 0,
\]
for the families Bell-diagonal state  under local dephasing.

\textbf{$\mathcal{N}_1 \leq \mathrm{ASC}$:}
For any non-negative real numbers $a,b,c$, one has $\max(a,b,c)\leq a+b+c$.
Since for our states
\[
\mathcal{N}_1(t)=\max\bigl(|c_1|,|c_2|,|c_3|\bigr),
\]
\[
\mathrm{ASC}(t)=|c_1|+|c_2|+|c_3|,
\]
the inequality follows immediately.

\textbf{$C(t) \leq \mathcal{N}_1(t)$:}
For X-states, concurrence is determined by competing off-diagonal coherences
with threshold subtractions of the form
$|\varrho_{14}| - (1-b_3)/4$ and $|\varrho_{23}| - (1+b_3)/4$.
Since positivity of the density matrix ensures $(1\pm b_3)\ge 0$ these
threshold terms only reduce the available coherences, implying
\[
C(t) \le 2\max\bigl(|\varrho_{14}|,|\varrho_{23}|\bigr).
\]

Using
\[
4|\varrho_{14}| = |b_1-b_2|e^{-2\kappa t}, \qquad
4|\varrho_{23}| = |b_1+b_2|e^{-2\kappa t},
\]
we obtain
\[
2\max(|\varrho_{14}|,|\varrho_{23}|)
= \tfrac{1}{2}\max(|b_1-b_2|,|b_1+b_2|)e^{-2\kappa t}.
\]

Using the inequality $|b_1\pm b_2|\le |b_1|+|b_2|$ and
$\max(a,b)\le a+b$ for $a,b\ge 0$, it follows that
\[
C(t) \le \max(|b_1|,|b_2|)e^{-2\kappa t}
\le \mathcal{N}_1(t).
\]

Combining both results, we obtain the ordering
\[
C(t) \leq \mathcal{N}_1(t) \leq \mathrm{ASC}(t),
\]
which holds for all $t \ge 0$ in the Bell-diagonal dephasing X-state family. We have checked this ordering numerically for the one-way steering state as well, using the exact solution and general definitions of Sec.~\ref{sec:oneway_deriv}, and it continues to hold there over the parameter range studied; we do not have an algebraic proof for that state since $\mathcal N_1$ and ASC are not available in closed form.

Thus concurrence is the least robust of the three measures under dephasing:
it decays fastest and can vanish at finite time, whereas $\mathcal N_1$ freezes at
$|b_3|$ and ASC approaches the same asymptote from above. The ordering concerns
the numerical values of the measures and is distinct from the resource-class
hierarchy of Refs.~\cite{Wiseman2007,Quintino2015}.

\section{Results and Discussion}
\label{sec:results}

We now present and analyze the numerical results for the time evolution of $C(t)$, $\mathcal{N}_1(t)$, and $\mathrm{ASC}(t)$ under pure dephasing for all the initial states. We work throughout in the dimensionless time variable $\Omega t$, where $\Omega = \alpha$ is the hyperfine coupling frequency, and quote dephasing strengths in units of $\Omega$. Restoring $\hbar$ explicitly, $\Omega=\alpha/\hbar\approx2.2\times10^{9}\,\mathrm{s^{-1}}$ for $\alpha\approx1.47\times10^{-6}$~eV (Sec.~\ref{sec:model}); the dephasing rates used below, $\kappa=0.02$--$0.5\,\alpha$, therefore correspond to $\kappa\approx4\times10^{7}$--$1\times10^{9}\,\mathrm{s^{-1}}$, consistent in order of magnitude with the coherence times reported for solid H$_2$ in Ref.~\cite{Sheludiakov2019}.

\subsection{Representative X-State}
\label{sec:results_hyperfine}
\begin{figure}[h]
    \centering
    \includegraphics[width=0.95\columnwidth]{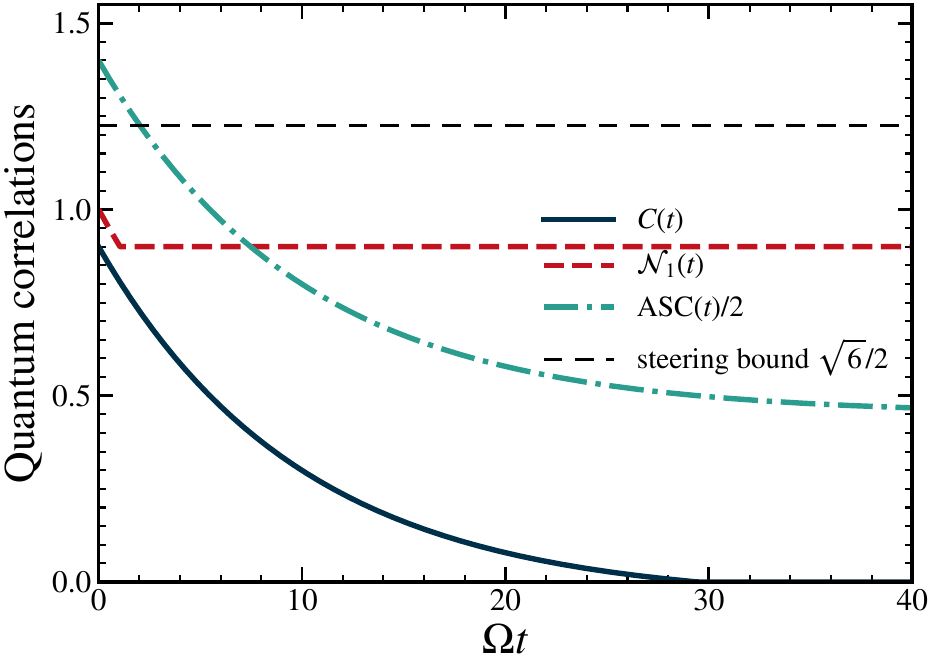}
    \caption{Concurrence $C$, Trace MIN $\mathcal{N}_1$ and $\mathrm{ASC}/2$ for
    $b_1=1$, $b_2=-0.9$, $b_3=0.9$ at $\kappa=0.05\,\alpha$. The dashed line is the
    steering bound $\sqrt6/2\approx1.225$. Steering is lost at
    $\Omega t=2.04$, entanglement at $\Omega t=29.4$.}
    \label{fig:all3_fixed_kappa}
\end{figure}

Although the physical ground state of the hydrogen hyperfine system is the maximally entangled singlet $b_1=b_2=b_3=-1$), for which Trace MIN  freezes trivially at $\mathcal{N}_1(t)=1$ with no visible crossover, we illustrate the full four-regime hierarchy using the representative X-state  $b_1=1$, $b_2=-0.9$, $b_3=0.9$. This state belongs to the physical  X-state family and is chosen because it makes all four dynamical regimes simultaneously visible. Figure~\ref{fig:all3_fixed_kappa} shows the time evolution of concurrence $C(t)$, $\mathcal{N}_1(t)$, and $\mathrm{ASC}(t)/2$ for $b_1=1$, $b_2=-0.9$, $b_3=0.9$ at $\kappa=0.05\,\alpha$. ASC is plotted as $\mathrm{ASC}/2$ for graphical comparison, with the steering threshold shown correspondingly at $\sqrt6/2$. Concurrence decays fastest and exhibits entanglement sudden death at a finite time $t^*$. Trace MIN remains finite beyond this point and saturates to $|b_3|$ at long times. In comparison, ASC decays more gradually, remaining nonzero after entanglement death but eventually approaching the same asymptotic scale. The horizontal dashed line at  $\mathrm{ASC}/2 = \sqrt6/2\approx1.225$ marks the  steering threshold~\cite{Mondal2017}: states with 
$\mathrm{ASC}/2 > \sqrt6/2$ exhibit EPR steering,  while states with $\mathrm{ASC}/2 \leq \sqrt6/2$ are non-steerable  under this witness, though they may still possess entanglement and discord-like correlations.

\begin{figure}[h]
    \centering
    \includegraphics[width=\columnwidth]{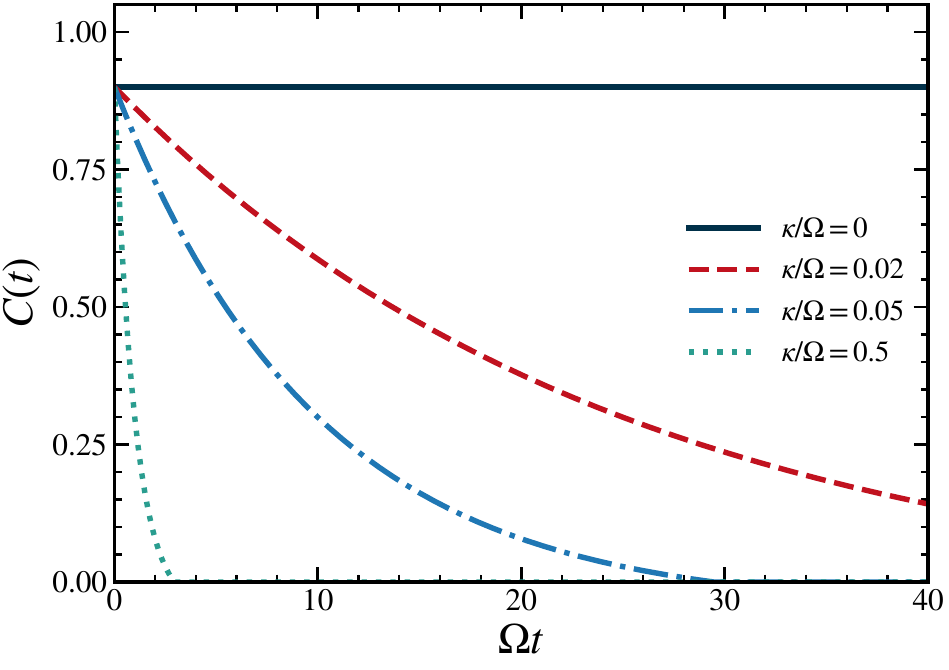}
    \caption{Concurrence for $b_1=1$, $b_2=-0.9$, $b_3=0.9$ at
    $\kappa=0$, $0.02\alpha$, $0.05\alpha$ and $0.5\alpha$. Stronger dephasing
    kills entanglement sooner.}
    \label{fig:concurrence_hyp}
\end{figure}

Figure~\ref{fig:concurrence_hyp} shows concurrence for the same state under different dephasing strengths $\kappa$. In the unitary limit ($\kappa=0$), entanglement is preserved. For $\kappa>0$, concurrence decays and the entanglement sudden death time scales approximately as $t^*\propto \kappa^{-1}$.

\begin{figure}[h]
    \centering
    \includegraphics[width=\columnwidth]{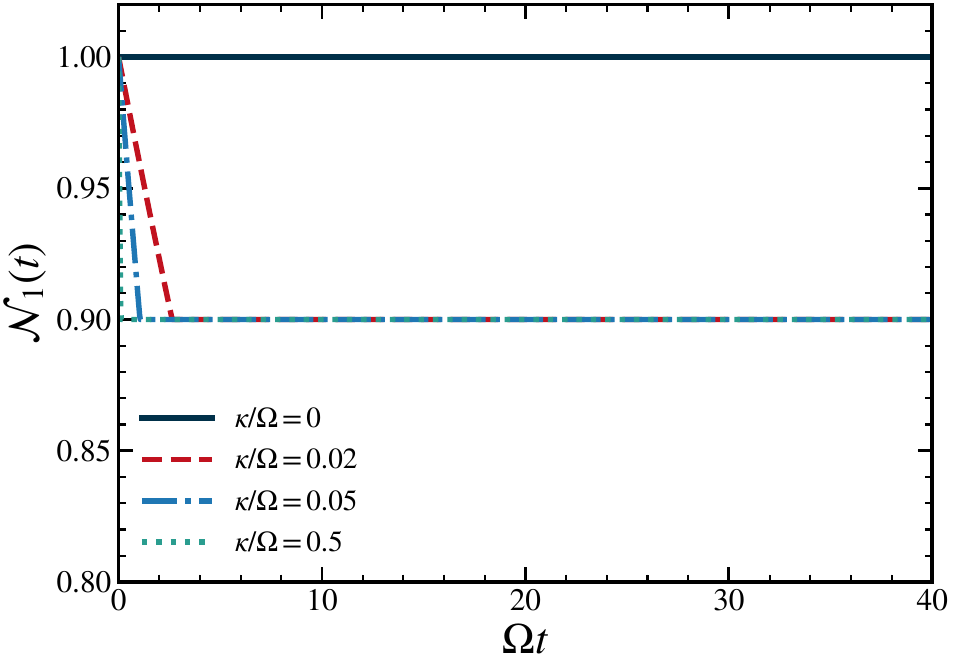}
    \caption{Trace MIN for the same state at four dephasing rates. Every curve
    freezes at $|b_3|$, whatever $\kappa$.Here $|b_1|>|b_3|$, so a real
    decay precedes the plateau.}
    \label{fig:tracemin_hyp}
\end{figure}

Figure~\ref{fig:tracemin_hyp} shows Trace MIN for the same parameter space. After a crossover time
\[
t^\dagger = \frac{1}{2\kappa}\ln\!\left(\frac{\max(|b_1|,|b_2|)}{|b_3|}\right),
\]
the dynamics freezes and $\mathcal{N}_1(t)$ saturates at $|b_3|$. This long-time plateau is independent of $\kappa$, reflecting the standard discord-freezing behavior.

\begin{figure}[h]
    \centering
    \includegraphics[width=\columnwidth]{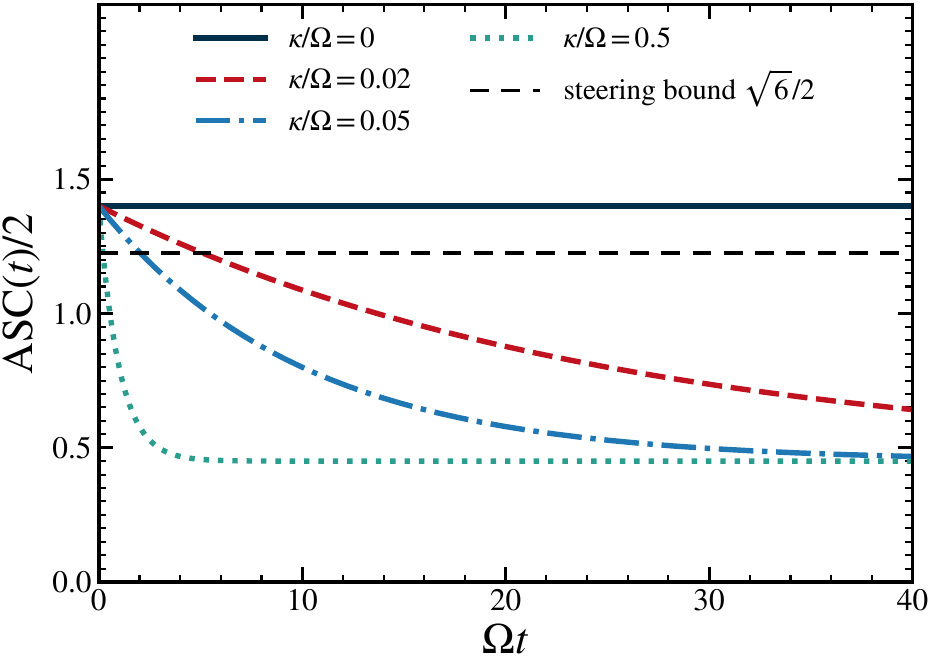}
    \caption{$\mathrm{ASC}/2$ for $b_1=1$, $b_2=-0.9$, $b_3=0.9$. The dashed
    line is the steering bound $\sqrt6/2\approx1.225$. }
    \label{fig:ASC_hyp}
\end{figure}

Figure~\ref{fig:ASC_hyp} presents $\mathrm{ASC}(t)/2$ for the same state. The horizontal line at $\sqrt6/2$ marks the steerability threshold. The state is initially steerable, but steering is lost at a finite time, earlier than entanglement sudden death. Unlike Trace MIN, ASC does not exhibit a sharp freezing crossover; it decays more gradually but approaches the same asymptote $|b_3|$ from above.

\subsubsection{Four dynamical regimes}
\label{sec:regimes}

\begin{figure}[h]
    \centering
    \includegraphics[width=\columnwidth]{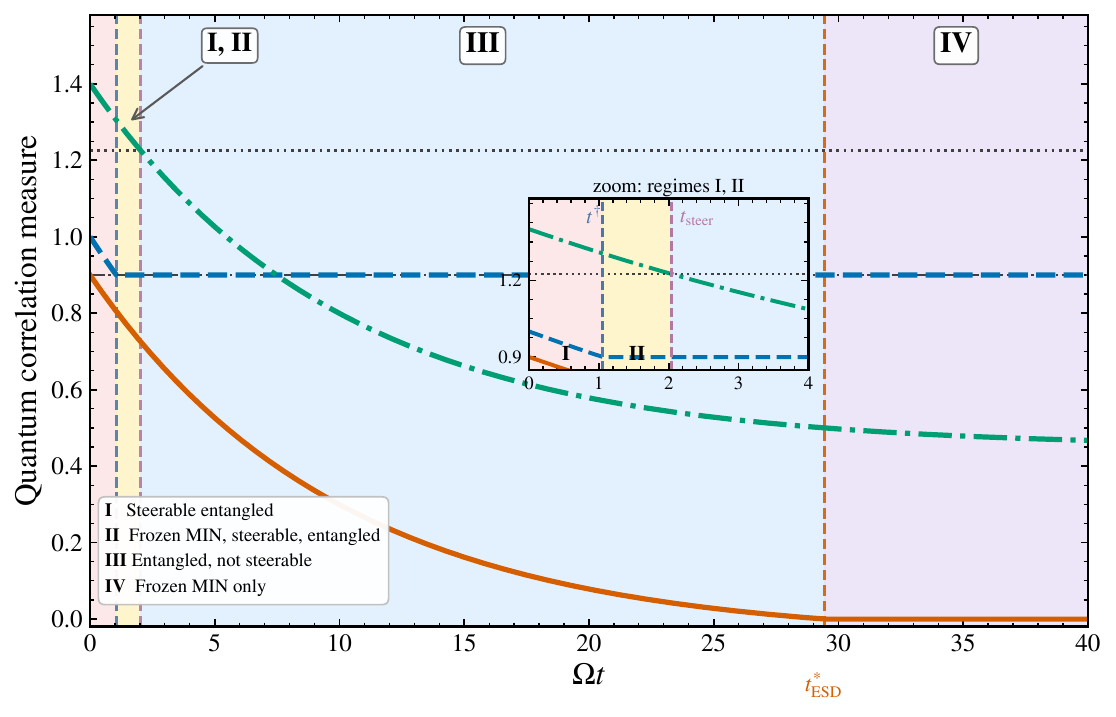}
    \caption{Four dynamical regimes for $b_1=1$, $b_2=-0.9$, $b_3=0.9$ at
    $\kappa=0.05\,\alpha$. Solid: $C$. Dashed: $\mathcal{N}_1$. Dash-dotted:
    $\mathrm{ASC}/2$. Two horizontal lines, not interchangeable: dotted is the
    steering bound $\sqrt6/2$, dash-dotted is $|b_3|=0.9$, the common
    asymptote. Vertical lines at $\Omega t^{\dagger}=1.05$ (freezing),
    $\Omega t_{\rm steer}=2.04$ and $\Omega t^{*}_{\rm ESD}=29.4$ separate the
    regimes: (I) steerable and entangled, (II) $\mathcal{N}_1$ frozen but still
    steerable and entangled, (III) entangled, no longer steerable, (IV)
    $\mathcal{N}_1$ frozen alone.}
    \label{fig:four_regimes}
\end{figure}

Figure~\ref{fig:four_regimes} summarizes the hierarchy of quantum
correlations during the dephasing dynamics for states with
$b_3\neq0$. The evolution can be divided into four distinct regimes,
characterized by the successive loss of steering and entanglement,
while Trace MIN remains frozen at $\mathcal{N}_1=|b_3|$.

\begin{enumerate}

\item \textbf{Steerable entangled regime} ($\mathrm{ASC}/2>\sqrt6/2$)

Here $C>0$, $\mathcal N_1$ is still decaying, and
$\mathrm{ASC}>\sqrt6$, so the state is entangled and is detected as steerable.

\item \textbf{Frozen MIN entangled regime}
($\mathrm{ASC}/2>\sqrt6/2$, $\mathcal{N}_1=|b_3|$, $C>0$)

After $t^{\dagger}=(2\kappa)^{-1}\ln[\max(|b_1|,|b_2|)/|b_3|]$, Trace MIN
reaches $|b_3|$ and stays there. Concurrence and the steering witness are both
still above threshold.

\item \textbf{Non-steerable entangled regime}
($\mathrm{ASC}/2\leq\sqrt6/2$, $C>0$)

At $t_{\mathrm{steer}}$, ASC drops below $\sqrt6$ and steering is no longer
detected, while $C>0$ and $\mathcal N_1=|b_3|$. Entanglement therefore outlives
the steering signature.

\item \textbf{Frozen discord regime}
($C=0$, $\mathcal{N}_1=|b_3|$)
At $t_{\mathrm{ESD}}$~\cite{Yu2004,Yu2009,Almeida2007} the concurrence
vanishes and the state becomes separable, leaving only the frozen
$\mathcal N_1=|b_3|$, which is set by the dephasing-invariant correlator
$\langle\sigma_z\otimes\sigma_z\rangle$.
\end{enumerate}
\subsection{Werner States}
\label{sec:results_werner}

\begin{figure}[h]
\centering
\includegraphics[width=\columnwidth]{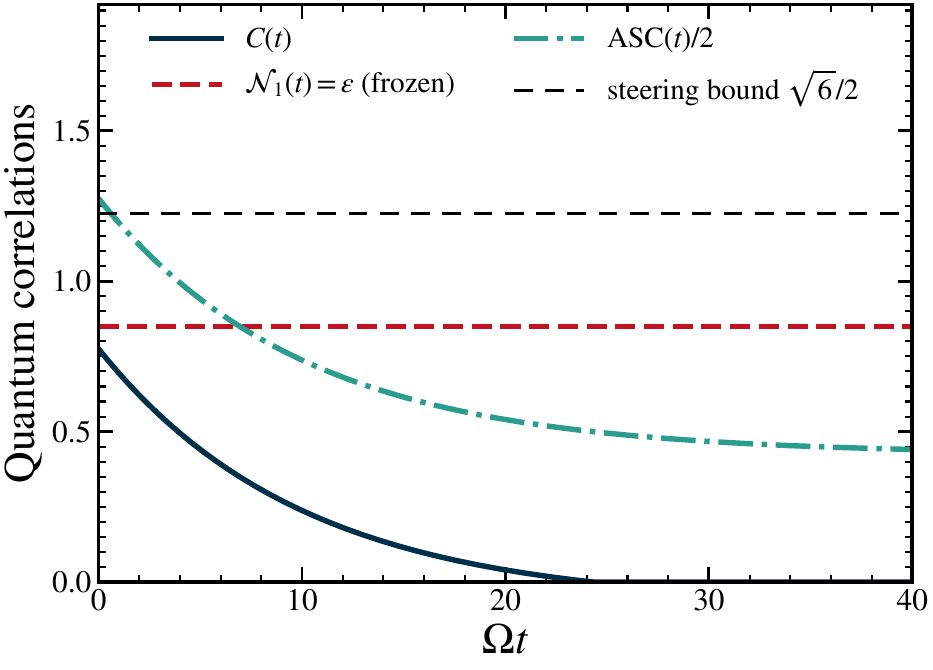}
\caption{Werner state $\varrho_+$ at $\varepsilon=0.85$, $\kappa=0.05\,\alpha$.
    Entanglement dies at finite time, while $\mathcal{N}_1=\varepsilon$
    from $t=0$: with $|b_1|=|b_2|=|b_3|=\varepsilon$ there is no decay phase at
    all.}
\label{fig:werner_baseline}
\end{figure}

Figure~\ref{fig:werner_baseline} shows the dynamical behavior of the three correlation measures for the Werner state $\varrho_+$ with $\varepsilon = 0.85$ under weak dephasing ($\kappa = 0.05\,\alpha$). The three quantities exhibit clearly different levels of robustness against environmental noise. Concurrence decreases rapidly and disappears completely after a finite interval, signaling entanglement sudden death. Because $|b_1|=|b_2|=|b_3|=\varepsilon$ for this state, Eq.~\eqref{eq:TMINfinal_clean} gives $\mathcal N_1(t)=\max(\varepsilon e^{-2\kappa t},\varepsilon e^{-2\kappa t},\varepsilon)=\varepsilon$ for all $t\ge0$: Trace MIN is constant at its $t=0$ value throughout, with no separate decay phase. ASC decays more gradually and remains the largest quantity throughout the evolution.

This behavior reflects the hierarchical structure of quantum correlations in the system. While entanglement is highly sensitive to decoherence, discord-type correlations encoded in Trace MIN survive much longer because the population correlation parameter $b_3$ is unaffected by pure dephasing. Consequently, the system enters a time window where nonclassical correlations persist despite the absence of entanglement.

\begin{figure*}[tbh]
\centering \includegraphics[width=0.32\textwidth]{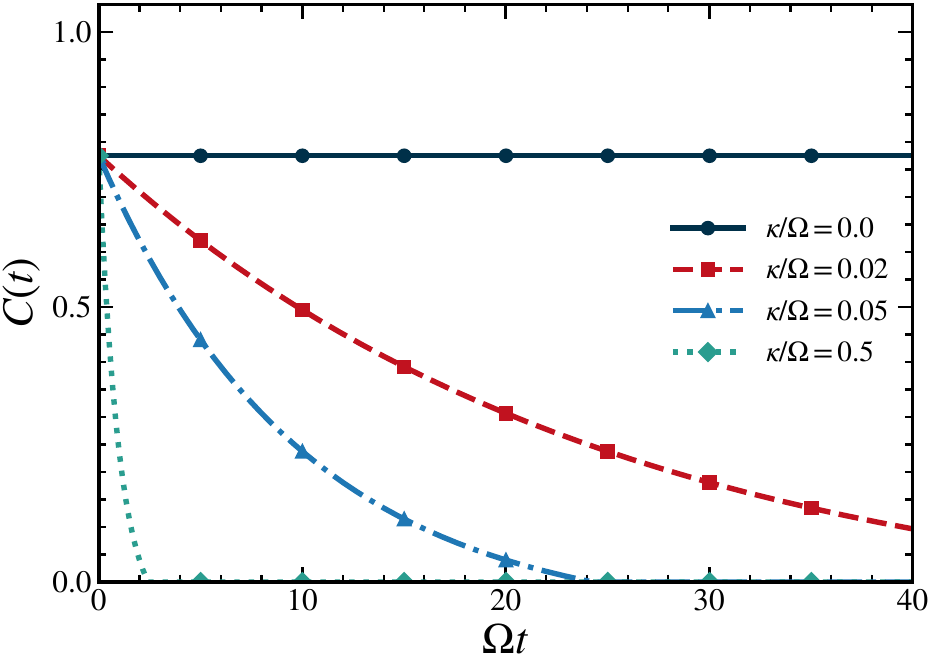}
\centering \includegraphics[width=0.32\textwidth]{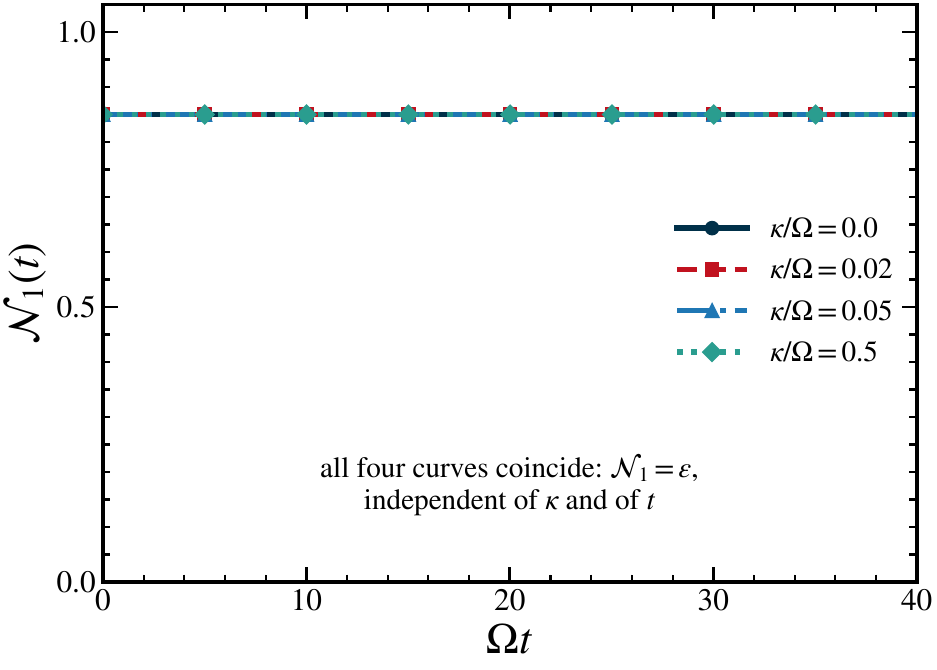}
\centering\includegraphics[width=0.32\textwidth]{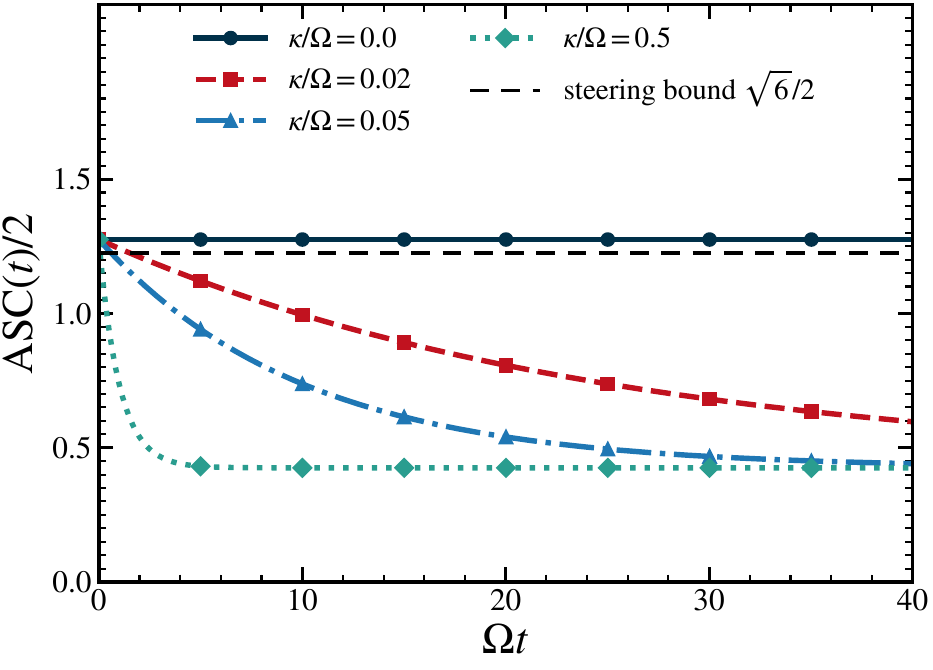}
\caption{(a) Concurrence, (b) Trace MIN and (c) $\mathrm{ASC}/2$ at four
    dephasing rates, $\varepsilon=0.85$. Panel (b) shows one visible
    line: $\mathcal{N}_1=\varepsilon$ at every $t$ and every $\kappa$, so the
    four curves coincide exactly. The dashed line in (c) is the steering bound
    $\sqrt6/2$.}
\label{fig:werner_kappa}
\end{figure*}

The influence of the dephasing strength is illustrated in Fig.~\ref{fig:werner_kappa} for fixed purity $\varepsilon = 0.85$. Increasing $\kappa$ accelerates the decay of the off-diagonal coherences, causing both concurrence and ASC to decrease more rapidly. The onset of entanglement sudden death therefore shifts to earlier times as the environment becomes more noisy.

A notable feature, however, is that the long-time value of Trace MIN remains unchanged for all dephasing rates considered(indeed, for this state it is constant at $\varepsilon$ from $t=0$, as noted above, so it is unaffected by $\kappa$ at every time, not only asymptotically). This occurs because the frozen contribution originates entirely from the constant population imbalance $b_3$, which is immune to phase damping. Dephasing modifies only the coherence terms $b_1$ and $b_2$, while leaving the asymptotic discord-like correlations intact.

\begin{figure*}[tbh]
\centering\includegraphics[width=0.32\textwidth]{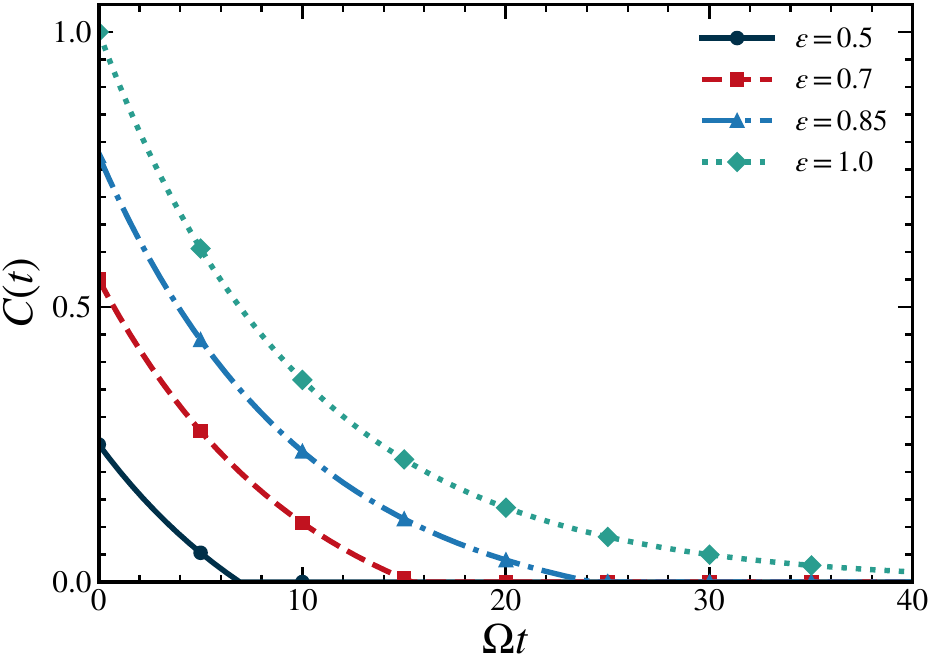}
\centering\includegraphics[width=0.32\textwidth]{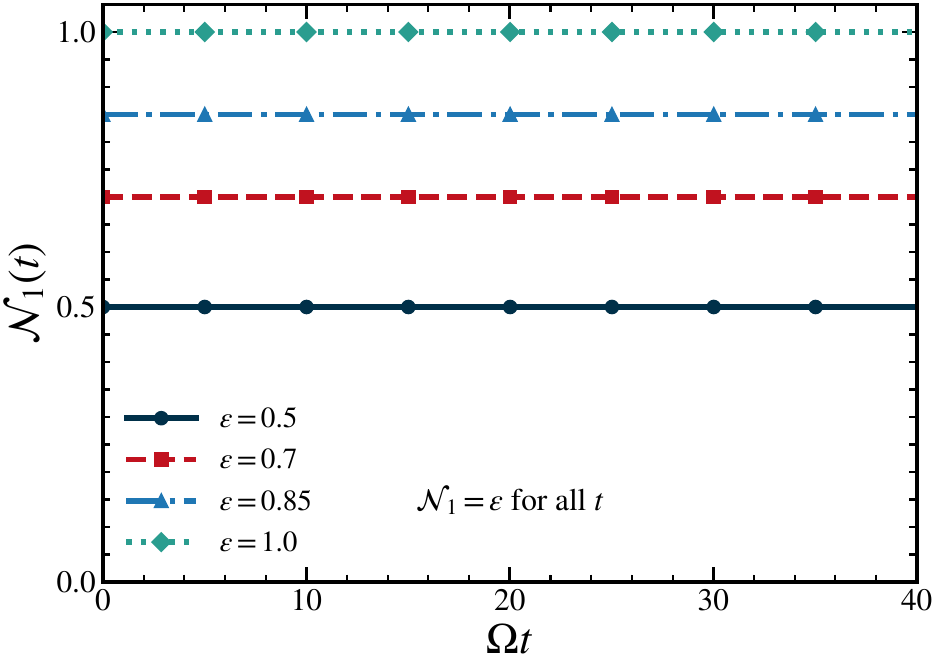}
\centering\includegraphics[width=0.32\textwidth]{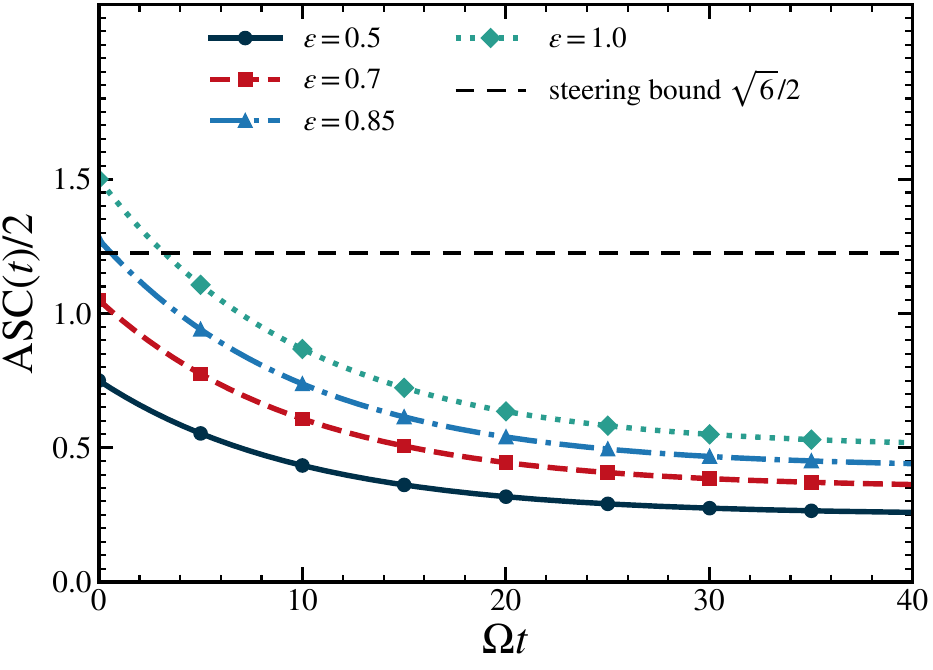}
\caption{(a) Concurrence, (b) Trace MIN and (c) $\mathrm{ASC}/2$ against the
    purity $\varepsilon$ at $\kappa=0.05\,\alpha$. Each curve in (b) is
    flat at its own $\varepsilon$. The dashed line in (c) is the steering bound
    $\sqrt6/2$, which only $\varepsilon\gtrsim0.82$ clears.}
\label{fig:werner_eps}
\end{figure*}

Figure~\ref{fig:werner_eps} shows the role of the Werner-state purity parameter $\varepsilon$ at fixed dephasing strength $\kappa = 0.05\,\alpha$. Increasing $\varepsilon$ enhances the initial amount of all three correlations and simultaneously raises the long-time saturation value of Trace MIN and ASC. Physically, larger $\varepsilon$ corresponds to a state closer to a maximally entangled Bell state, leading to stronger nonclassical correlations at all times.

For smaller $\varepsilon$, the correlations decay more rapidly and the frozen plateau becomes less pronounced. Nevertheless, as long as $\varepsilon \neq 0$, the nonzero parameter $b_3=\varepsilon$ guarantees the persistence of a finite asymptotic value for Trace MIN. The freezing is therefore set by the population sector of the density matrix, not by the decaying coherences.

\subsection{One-Way Steering State}
\label{sec:results_steering}
As established in Sec.~\ref{sec:oneway_deriv}, this state's marginals are not maximally mixed for $\theta\ne\pi/4$, and its dynamics is governed by the full Lindblad equation with the Hamiltonian term retained, not by the Bell-diagonal solution Eqs.~\eqref{eq:rho11}--\eqref{eq:rho14}. Everything below for $p=0.85,\theta=0.4$ is computed from the exact solution of Sec.~\ref{sec:oneway_deriv}: $C(t)$ from Eq.~\eqref{eq:sw_concurrence} in closed form, and $\mathcal N_1(t)$, $\mathrm{ASC}(t)$ by direct numerical evaluation of Eqs.~\eqref{eq:TMINdef} and~\eqref{eq:ASC_def} on $\varrho(t)$.

The one-way steering state is characterized by $\langle ZZ\rangle=p$ exactly, for any $\theta$ (Eq.~\eqref{eq:sw_ZZ}), and $\langle XX\rangle=-\langle YY\rangle=p\sin2\theta\,e^{-2\kappa t}$ (Eqs.~\eqref{eq:sw_XX} and~\eqref{eq:sw_YY}). Unlike the Werner states, the density matrix is generally asymmetric for $\theta\neq \pi/4$: the population imbalance $w(t)=\varrho_{22}(t)-\varrho_{33}(t)$ executes a damped oscillation at frequency $\nu\approx4\Omega$, decaying at rate $\kappa$ rather than $2\kappa$. This oscillation is driven by the hyperfine coupling $\Omega$ itself and has no counterpart in the Bell-diagonal states discussed above, where $\Omega$ never enters the dynamics. The long-time behavior, however, is not governed by the same mechanism as in the Bell-diagonal case, and this is a direct consequence of the marginals. Because $\varrho_A\ne\mathbb I/2$ for $\theta\ne\pi/4$, the only admissible Trace~MIN measurement is the one along the Bloch vector of $\varrho_A$ (Sec.~\ref{sec:tracemin_def}), here the $\hat z$ direction, and that measurement is sensitive only to the off-block coherences $\varrho_{14}$ and $\varrho_{23}$. Both are damped, so $\mathcal N_1(t)\to0$: the freezing at $|b_3|$ found for the Bell-diagonal family does not occur here. It is recovered only in the limit $\theta=\pi/4$, where $\varrho_A$ becomes maximally mixed, every direction becomes admissible, and $\mathcal N_1\to p$. The constancy of $\langle ZZ\rangle=p$ is therefore not by itself sufficient for freezing; a maximally mixed marginal is required as well.

\begin{figure}[!ht]
\centering
\includegraphics[width=\columnwidth]{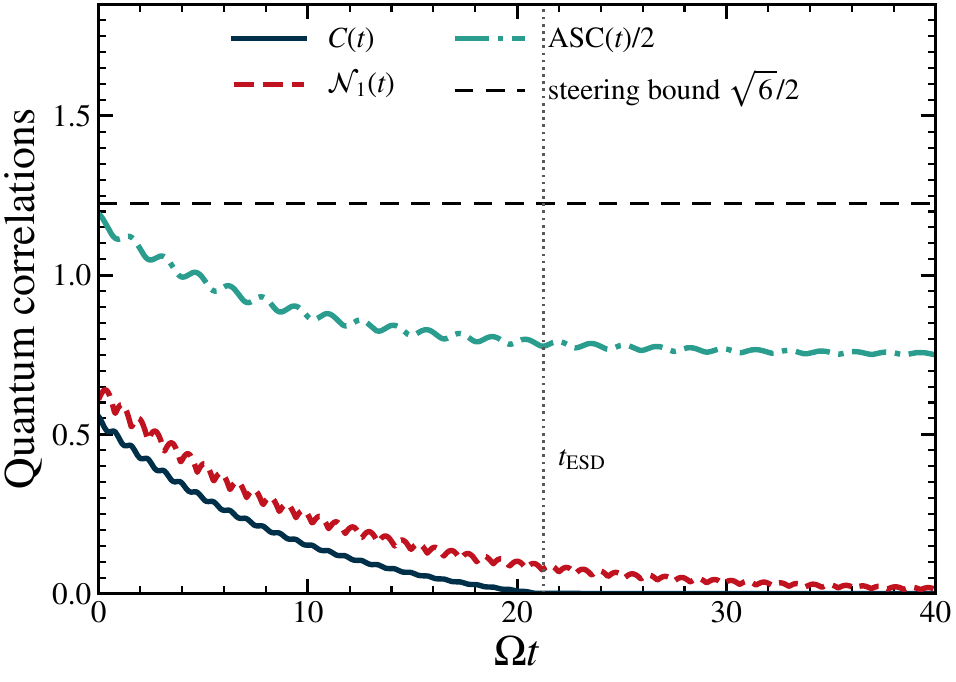}
\caption{One-way steering state at $p=0.85$, $\theta=0.4$,
    $\kappa=0.05\,\alpha$, from the exact solution with the Hamiltonian
    term kept; $\mathcal{N}_1$ and $\mathrm{ASC}$ are evaluated numerically.
    }
\label{fig:steering_baseline}
\end{figure}

Under phase damping, the coherence-dependent contributions decay. Concurrence disappears at a finite time ($\Omega t_{\rm ESD}=21.26$ for the parameters below), and $\mathcal N_1$ decays to zero with it, for the reason just given. $\mathrm{ASC}$ does approach a nonzero constant, but that residual value originates in the $\ell_1$ coherence of Bob's own marginal rather than in any correlation between the qubits, and it lies far below the steering bound; it should not be read as a surviving nonclassical correlation. Figure~\ref{fig:steering_baseline} illustrates the evolution of the three correlation measures for $p=0.85$, $\kappa = 0.05\,\alpha$, and $\theta=0.4$. Initially, the state possesses moderate entanglement together with stronger discord-like and steering-type correlations. At $t=0$ the local Bloch components Eq.~\eqref{eq:sw_local_bloch} are nonzero, so $\mathcal N_1(0)=0.610$ and $\mathrm{ASC}(0)=2.386$ are obtained from the operational definitions Eqs.~\eqref{eq:TMINdef} and \eqref{eq:ASC_def}  The hierarchy $
[
C(t)\leq \mathcal{N}_1(t)\leq \mathrm{ASC}(t)
] $
continues to hold (checked numerically, Sec.~\ref{sec:hierarchy_proof}) throughout the dynamics.

Concurrence decays most rapidly and undergoes entanglement sudden death, while $\mathcal{N}_1$ decays alongside it, carrying a visible ripple: $\varrho_{23}(t)\propto|\sin\nu t|$, so both $\mathcal N_1$ and ASC inherit a rectified oscillation at $\approx8\Omega$. The ripple is absent from the Bell-diagonal family. ASC remains the largest of the three measures, but at these parameters it does not detect steering at any time: $\mathrm{ASC}(0)=2.386$ lies below $\sqrt6=2.449$, although the state is one-way steerable according to Ref.~\cite{Bowles2014,Bowles2016}. The bound is crossed for larger mixing, $\mathrm{ASC}(0)=2.606$ at $p=0.95$ and $2.717$ at $p=1$.

\begin{figure*}[tbh]%
\centering \includegraphics[width=0.32\textwidth]{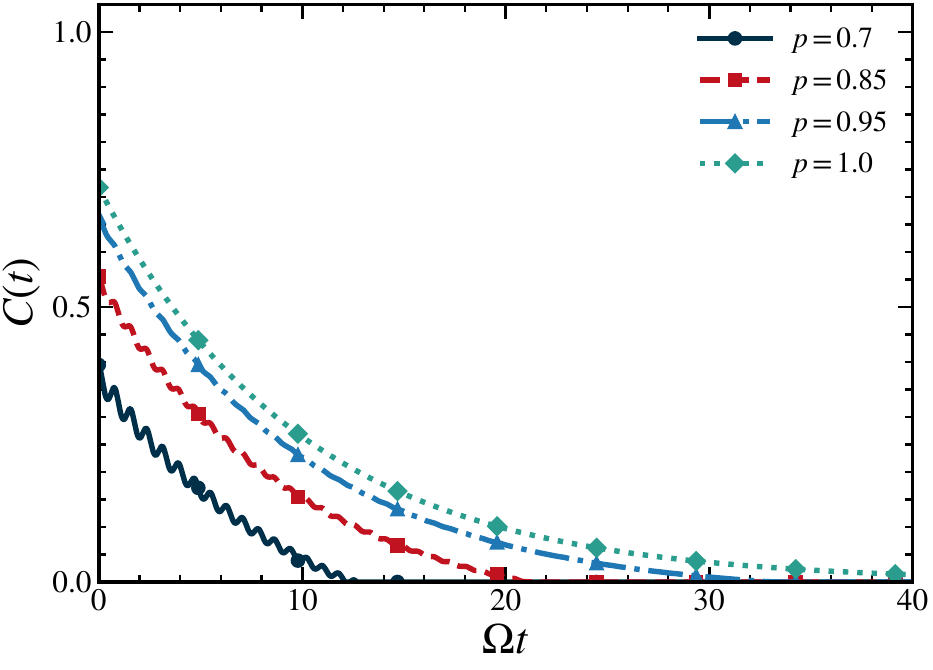}
\centering \includegraphics[width=0.32\textwidth]{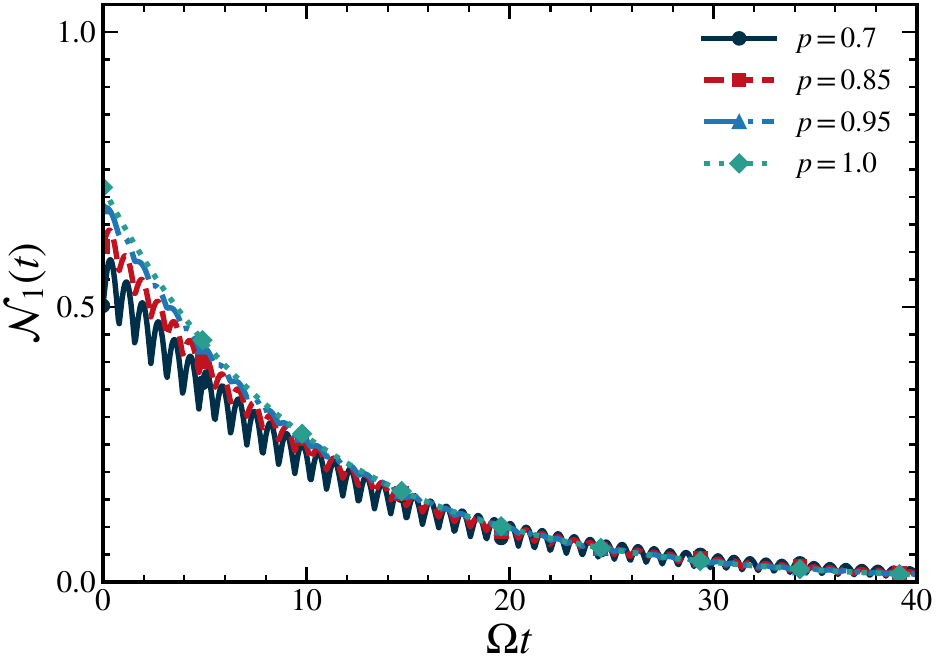}
\centering\includegraphics[width=0.32\textwidth]{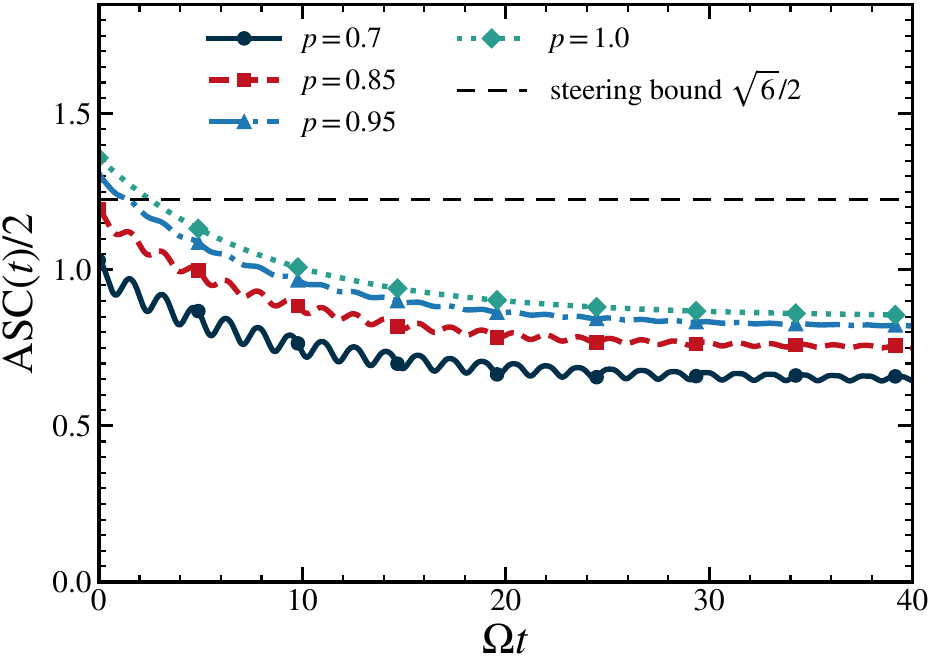}
\caption{Effect of the mixture parameter $p$ on (a) concurrence, (b) Trace MIN
    and (c) $\mathrm{ASC}/2$ at $\kappa=0.05\,\alpha$, $\theta=0.4$.
    The dashed line in (c) is the steering bound $\sqrt6/2$}
\label{fig:steering_vary_p}
\end{figure*}

Figures~\ref{fig:steering_vary_p}(a)--(c) show the effect of varying the mixture parameter $p$ while keeping $\kappa = 0.05\,\alpha$ and $\theta=0.4$ fixed. Increasing $p$ enhances the initial values of all three correlation measures, indicating a larger contribution from the coherent entangled component of the state.

Larger $p$ also leads to a higher long-time saturation value for both $\mathcal{N}_1$ and ASC, since the asymptotic correlations are directly proportional to the conserved quantity $\langle ZZ\rangle=p$. Consequently, states with larger mixture weight retain stronger nonclassical correlations even after significant decoherence has occurred.

\begin{figure*}[tbh]%
\centering\includegraphics[width=0.32\textwidth]{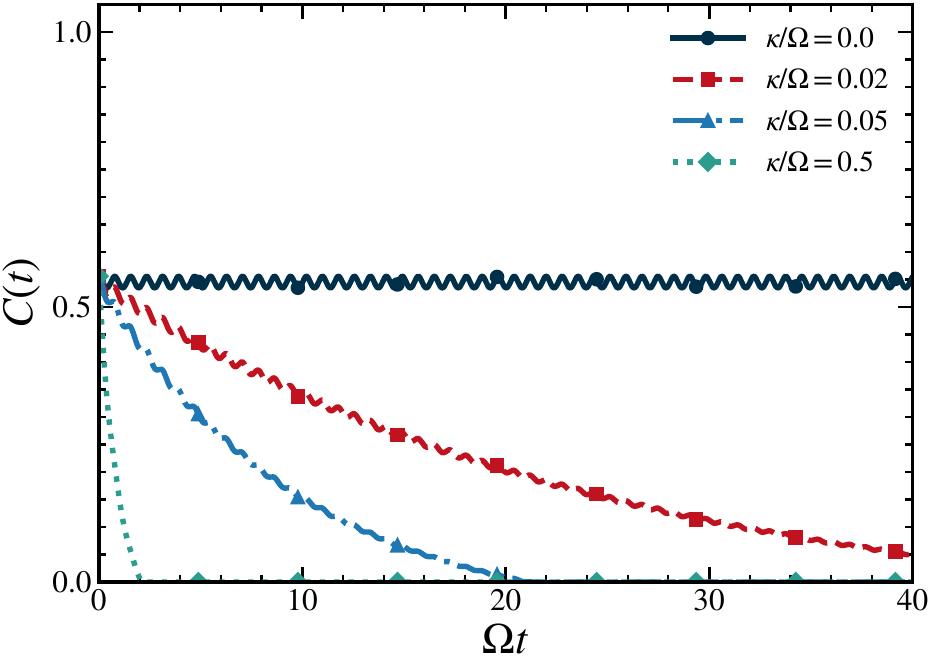}
\centering\includegraphics[width=0.32\textwidth]{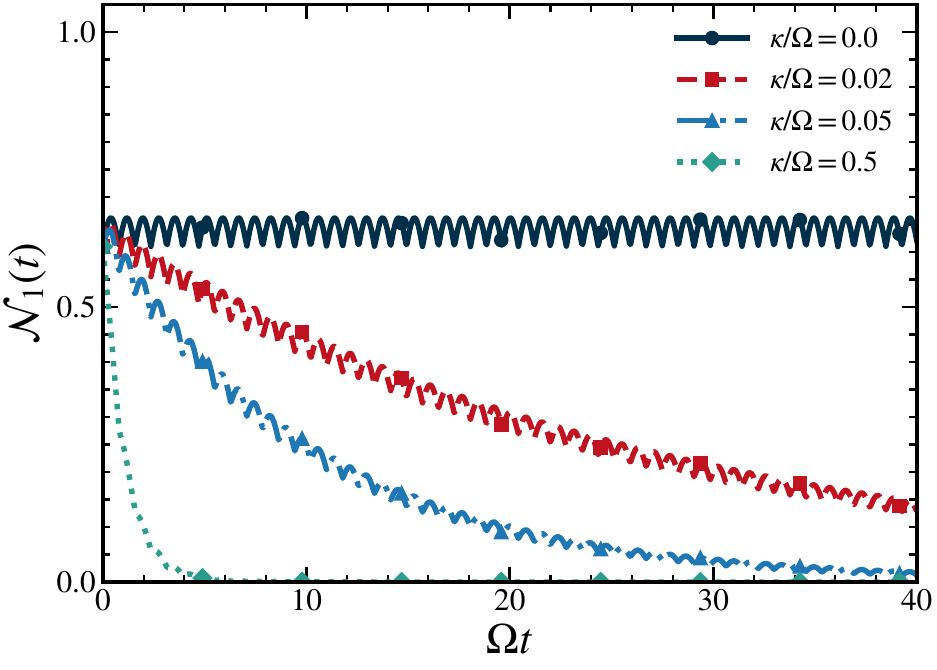}
\centering\includegraphics[width=0.32\textwidth]{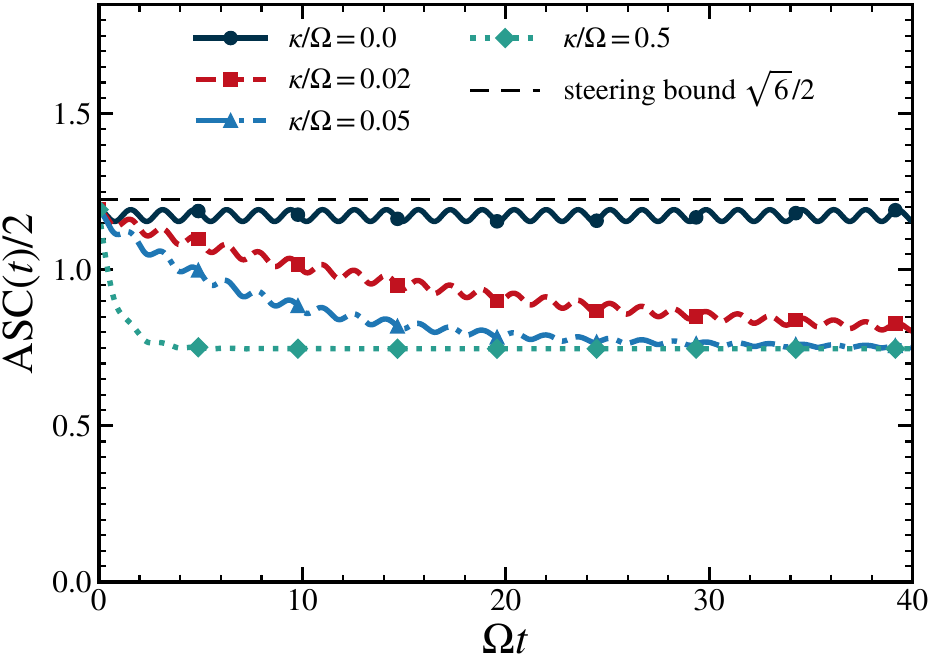}
\caption{(a) Concurrence, (b) Trace MIN and (c) $\mathrm{ASC}/2$ at four
    dephasing rates, $p=0.85$, $\theta=0.4$. The dashed line in (c) is
    the steering bound $\sqrt6/2$.}
\label{fig:steering_vary_kappa}
\end{figure*}

Figures~\ref{fig:steering_vary_kappa}(a)--(c) present the effect of the dephasing strength for fixed $p=0.85$ and $\theta=0.4$. In the absence of decoherence $(\kappa=0)$, the $\{|01\rangle,|10\rangle\}$ block continues to oscillate indefinitely at frequency $4\Omega$  with $\kappa=0$) rather than remaining static, since the Hamiltonian alone still drives this state; only the $\{|00\rangle,|11\rangle\}$ block is static. This is a genuine difference from the Bell-diagonal states, for which $\kappa=0$ gives a fully static state throughout. Once dephasing is introduced, the decay of the coherence-dependent terms becomes progressively faster with increasing $\kappa$, and for $\kappa\gtrsim4\Omega$ the oscillation in $w(t)$ would become overdamped. That regime lies well outside the range $\kappa=0.02$--$0.5\,\alpha$ motivated in Sec.~\ref{sec:results} and is noted here only as a property of the solution; the figures stay within the physical range.

Although stronger dephasing accelerates the suppression of concurrence and shortens the steerable time interval, the asymptotic values of $\mathcal{N}_1$ and ASC remain unchanged. This again highlights the central role of the conserved population correlator $\langle ZZ\rangle=p$, which is insensitive to phase noise and therefore determines the residual long-time correlations.

\section{Quantum Teleportation via the Dephased Hyperfine Channel}
\label{sec:teleportation}

The correlation hierarchy discussed in Sec.~\ref{sec:measures}
naturally raises an operational question: which of these correlations
remain useful for a concrete quantum-information task?
To address this, we employ the time-evolved thermal hyperfine state
$\varrho(T,t)$ as a noisy teleportation channel and evaluate its
average teleportation fidelity using the Horodecki
criterion~\cite{Horodecki1999},\cite{Verstraete2003}.
For a general two-qubit state, entanglement does not imply $F_A>2/3$: the Horodecki criterion gives $F_A>2/3\iff f>1/2$, and entangled states with $f\le1/2$ exist~\cite{Horodecki1999}. For the Bell-diagonal X-state family considered here the maximally mixed marginals give $a+d=1/2$, and the conditions $f>1/2$ and $C>0$ then coincide exactly. This equivalence is specific to this family.

At thermal equilibrium the electron--proton spin pair is described by
the Gibbs state
\begin{equation}
    \varrho_{\rm th}(T)
    = \frac{e^{-\beta H}}{Z},
    \qquad
    Z = e^{3\beta\alpha} + 3e^{-\beta\alpha},
    \quad
    \beta = \frac{1}{k_B T}.
    \label{eq:Gibbs}
\end{equation}
Using the singlet energy $E_s=-3\alpha$ and triplet energy
$E_t=+\alpha$, the thermal state in the computational basis takes the
X-state form
\begin{equation}
    \varrho_{\rm th}(T) =
    \begin{pmatrix}
      a & 0 & 0 & 0 \\
      0 & d & c_0 & 0 \\
      0 & c_0 & d & 0 \\
      0 & 0 & 0 & a
    \end{pmatrix},
    \label{eq:rho_thermal}
\end{equation}
where
\begin{align}
    a   &= \frac{e^{-\beta\alpha}}{Z},
    \label{eq:a_th}\\
    d   &= \frac{e^{3\beta\alpha}+e^{-\beta\alpha}}{2Z},
    \label{eq:d_th}\\
    c_0 &= -\frac{e^{3\beta\alpha}-e^{-\beta\alpha}}{2Z} < 0.
    \label{eq:c0_th}
\end{align}
Note that $\varrho_{14}=0$ identically, so only the $\varrho_{23}$
coherence is active.  Under local pure dephasing, populations are
frozen while the coherence decays exponentially:
\begin{equation}
    c(t) = c_0\,e^{-2\kappa t},
    \label{eq:ct}
\end{equation}
giving the time-evolved channel $\varrho_{\rm th}(T,t)$ used as the
entangled resource for teleportation.

The average teleportation fidelity over all input states is related
to the fully entangled fraction~\cite{Horodecki1999}
\begin{equation}
    f(T,t)
    = \max_{|\Phi\rangle\in\text{Bell}}
      \langle\Phi|\,\varrho_{\rm th}(T,t)\,|\Phi\rangle,
    \label{eq:fef}
\end{equation}
by $F_A = (1+2f)/3$.  The four Bell-state overlaps for the
state~(\ref{eq:rho_thermal}) are
\begin{align}
    \langle\Phi^\pm|\varrho|\Phi^\pm\rangle &= a,
    \label{eq:overlap_phi}\\
    \langle\Psi^+|\varrho|\Psi^+\rangle     &= d + c(t),
    \label{eq:overlap_psiplus}\\
    \langle\Psi^-|\varrho|\Psi^-\rangle     &= d - c(t).
    \label{eq:overlap_psiminus}
\end{align}
Since $c(t)<0$ for all $t\geq0$ (Eq.~\ref{eq:c0_th}), the singlet
$|\Psi^-\rangle$ always yields the maximum overlap:
\begin{equation}
    f(T,t)
    = d - c(t)
    = \frac{e^{3\beta\alpha}+e^{-\beta\alpha}
             +\!\left(e^{3\beta\alpha}-e^{-\beta\alpha}\right)e^{-2\kappa t}}
           {2Z}.
    \label{eq:fef_thermal}
\end{equation}
The average teleportation fidelity is therefore
\begin{equation}
   {
    F_A(T,t)
    = \frac{1}{3}\!\left[1
      + \frac{e^{3\beta\alpha}+e^{-\beta\alpha}
              +\!\left(e^{3\beta\alpha}-e^{-\beta\alpha}\right)e^{-2\kappa t}}
             {Z}
      \right].}
    \label{eq:FA_thermal}
\end{equation}

At long times ($t\rightarrow\infty$), the coherence vanishes,
$c(t)\rightarrow0$, and the fully entangled fraction reduces to
\begin{equation}
    f(T,\infty) = d
    = \frac{e^{3\beta\alpha}+e^{-\beta\alpha}}{2Z}.
\end{equation}
The average fidelity then becomes
\begin{equation}
    F_A(T,\infty)
    = \frac{1+2d}{3},
\end{equation}

At $t=0$, $F_A(T,0)\to1$ as $T\to0$, corresponding to the pure
singlet channel.

The classical teleportation threshold $F_A>2/3$ is equivalent to
$f>1/2$.  Using the identity $a+d=1/2$ (which follows from
$\mathrm{Tr}[\varrho_{\rm th}]=1$ and the symmetric structure
$\varrho_{11}=\varrho_{44}=a$, $\varrho_{22}=\varrho_{33}=d$), the
condition $d+|c(t)|>1/2=a+d$ simplifies directly to
\begin{equation}
    f > \tfrac{1}{2}
    \quad\Longleftrightarrow\quad
    |c(t)| > a
    \quad\Longleftrightarrow\quad
    C(T,t) > 0,
    \label{eq:FA_C_equivalence}
\end{equation}
as the concurrence formula $C = 2\max(0,|c|-a)$.
Solving $|c(t_F)|=a$ explicitly gives
\begin{equation}
   t_F(T) = t_{\rm ESD}(T)
    = \frac{1}{2\kappa}
      \ln\!\left(\frac{e^{4\beta\alpha}-1}{2}\right).
    \label{eq:tF_thermal}
\end{equation}

Thus, for the thermal hyperfine channel, the loss of teleportation
advantage occurs exactly at the time when entanglement disappears.
The teleportation advantage window is $[0, t_{\rm ESD}]$.

The concurrence at arbitrary time is
\begin{equation}
    C(T,t)
    = \max\!\left[0,\;
      \frac{\left(e^{3\beta\alpha}-e^{-\beta\alpha}\right)e^{-2\kappa t}
            - 2e^{-\beta\alpha}}{Z}
      \right],
    \label{eq:C_thermal}
\end{equation}
and while $C(T,t)>0$ the fidelity satisfies
\begin{equation}
    F_A(T,t)
    = \frac{2}{3} + \frac{C(T,t)}{3},
    \qquad C(T,t) > 0.
    \label{eq:FA_C_relation}
\end{equation}
Once concurrence vanishes, the fidelity remains bounded by
$F_A\leq2/3$.

\subsection{Generalization to All Three Initial States}
\label{sec:tel_allstates}

The equivalence $t_F=t_{\rm ESD}$ established above for the thermal
channel extends to the class of dephasing-driven X states considered
in this work, all of which possess maximally mixed marginals.
The key structural property is $a+d=1/2$, which holds for any X-state
in this family since
$\varrho_{11}+\varrho_{22}=(1+b_3)/4+(1-b_3)/4=1/2$.

\paragraph{Werner state $\varrho_-$.}
With $b_1=b_2=-\varepsilon$, $b_3=-\varepsilon$, only $\varrho_{23}$
is active and $f_- = d_- + |\varrho_{23}(t)|$ where
$d_-=(1+\varepsilon)/4$, $a_-=(1-\varepsilon)/4$.
The condition $f_->1/2$ gives $|\varrho_{23}|>a_-$, identical to
$C_->0$.  Hence $t_F=t_{\rm ESD}$.

\paragraph{Werner state $\varrho_+$.}
With $b_1=\varepsilon$, $b_2=-\varepsilon$, $b_3=\varepsilon$, only
$\varrho_{14}$ is active.
$f_+ = a_+ + |\varrho_{14}(t)|$ with $a_+=(1+\varepsilon)/4$,
$d_+=(1-\varepsilon)/4$.
The condition $f_+>1/2$ gives $|\varrho_{14}|>d_+$, identical to
$C_+>0$.  Hence $t_F=t_{\rm ESD}$.

\paragraph{One-way steering state $\varrho_\theta$ (at $\theta=\pi/4$)}.
The argument here uses $a+d=1/2$ together with the two-population structure $\varrho_{11}=\varrho_{44}=a$, $\varrho_{22}=\varrho_{33}=d$, which the one-way steering state has only at $\theta=\pi/4$ (Sec.~\ref{sec:xstate}); at general $\theta$, $\varrho_{11}\ne\varrho_{44}$ and a fully entangled fraction calculation would need to be carried out separately from the four-population state of Sec.~\ref{sec:oneway_deriv}, which we have not done here. This teleportation equivalence is therefore stated only at $\theta=\pi/4$ -- a narrower restriction than on the correlation measures $C$, $\mathcal N_1$, ASC themselves, which we now have for general $\theta$ (Sec.~\ref{sec:oneway_deriv}).
With $b_1=p\sin2\theta$, $b_2=-p\sin2\theta$, $b_3=p$,
one has $b_1+b_2=0$ so $\varrho_{23}=0$ and only $\varrho_{14}$ is
active.  With $a_s=(1+p)/4$, $d_s=(1-p)/4$,
$f_s=a_s+|\varrho_{14}(t)|$, and $f_s>1/2$ gives
$|\varrho_{14}|>d_s$, which is exactly $C_s>0$.
Hence $t_F=t_{\rm ESD}$ at $\theta=\pi/4$.

The general result for this family is
\begin{equation}
\begin{aligned}
    F_A > \tfrac{2}{3}\;&\Longleftrightarrow\;C > 0\\
    &\text{for all Bell-diagonal X-states}\\
    &\text{with }\varrho_A=\varrho_B=\tfrac{\mathbb{I}}{2}.
\end{aligned}
\label{eq:FA_C_general}
\end{equation}
This equivalence is specific to this maximally-mixed-marginal family and should not be read as applying to arbitrary two-qubit or X-states; the optimal (locally-filtered) teleportation fidelity for a general mixed two-qubit resource is treated in Ref.~\cite{Verstraete2003}.

Equations~(\ref{eq:FA_C_equivalence}) and~(\ref{eq:FA_C_general})
have a clear operational consequence.  The hierarchy
$C(t)\leq\mathcal{N}_1(t)\leq\mathrm{ASC}(t)$ established in
Sec.~\ref{sec:measures} shows that Trace MIN and ASC survive beyond
ESD: both approach the nonzero asymptote $|b_3|$ from above, while
$C$ has already vanished.  Yet by Eq.~(\ref{eq:FA_C_general}), these
residual correlations provide no advantage for standard quantum
teleportation: once $C=0$, $F_A\leq2/3$.

For the state family considered here,
\begin{equation}
\begin{aligned}
    \underbrace{C>0}_{\substack{\text{teleportation}\\\text{advantage}}}
    \;\Longrightarrow\;
    \underbrace{\mathcal{N}_1>0}_{\substack{\text{nonzero discord-}\\\text{type correlation}}}
    \;\Longrightarrow\;
    \underbrace{\mathrm{ASC}>0}_{\substack{\text{nonzero steering}\\text{coherence}}} ,
\end{aligned}
\label{eq:teleportation_hierarchy}
\end{equation}
so $\mathcal N_1$ and ASC can both persist after the concurrence vanishes,
but neither then supports a teleportation advantage. Nonzero ASC does not
certify steering either; the steering condition is $\mathrm{ASC}>\sqrt6$.

\begin{figure}[t]
    \centering
    \includegraphics[width=\columnwidth]{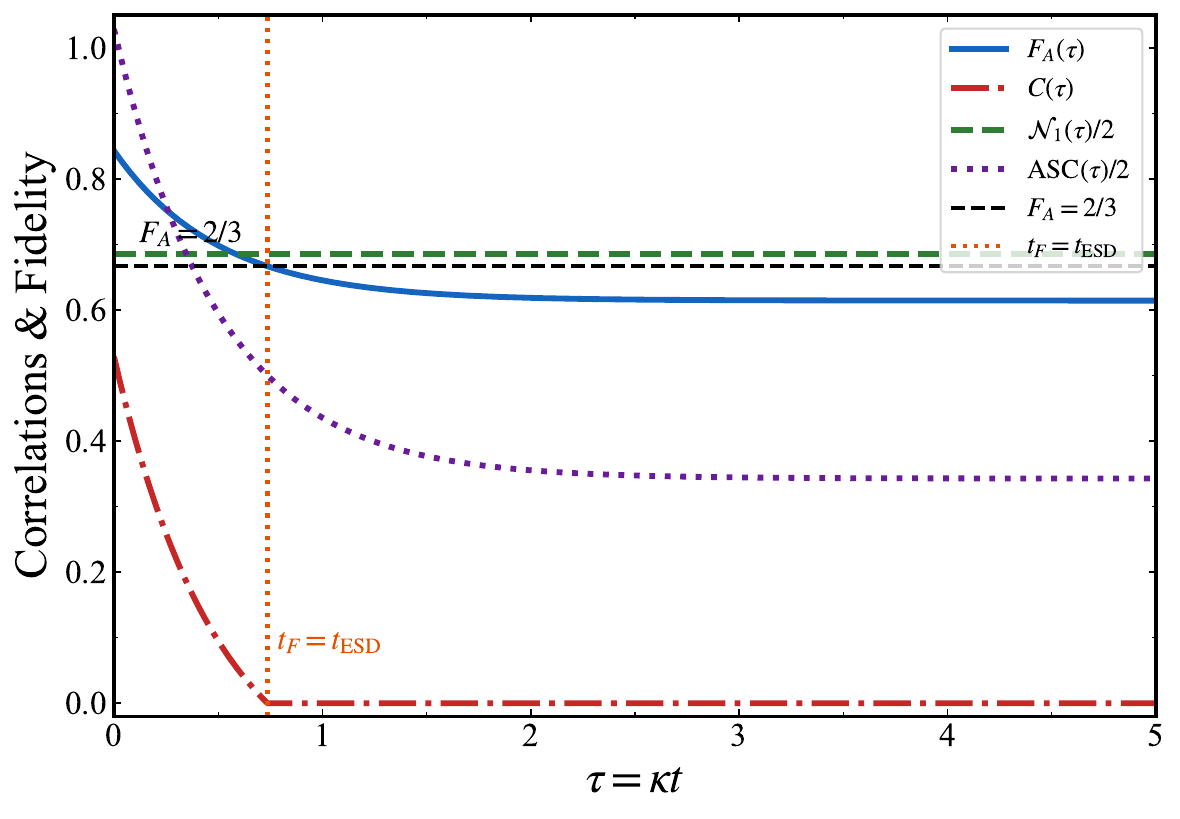}
    \caption{
Teleportation fidelity and correlation dynamics for the dephased thermal
hyperfine channel at $T=0.03$~K. The dashed horizontal line is the
classical bound $F_A=2/3$; the dotted vertical is the teleportation lifetime
$\tau_F$, which falls exactly at entanglement sudden death.}

    \label{fig:tel_thermal}
\end{figure}

\begin{figure}[t]
    \centering
    \includegraphics[width=\columnwidth]{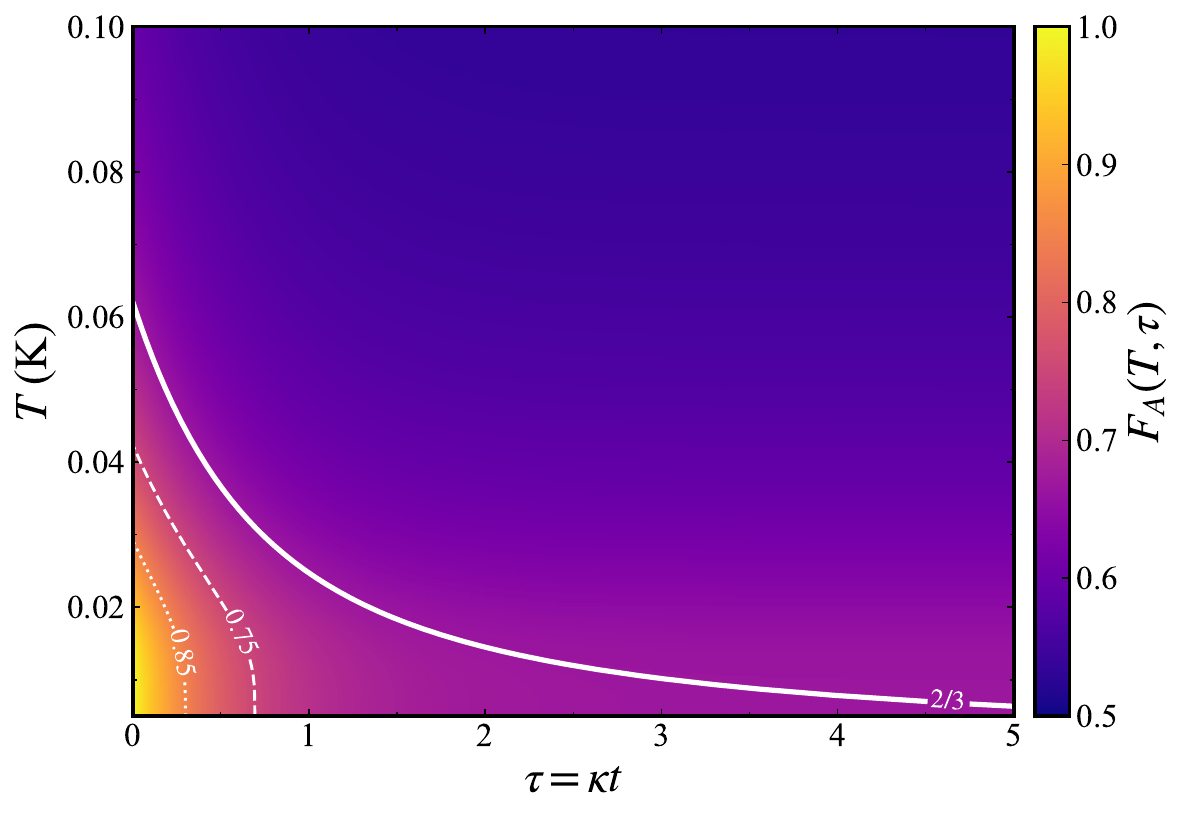}
    \caption{
Teleportation fidelity of the dephased thermal channel against
temperature and rescaled time $\tau=\kappa t$. The white contour is
$F_A=2/3$: below it teleportation beats any classical protocol, above it does
not.}

    \label{fig:tel_heatmap}
\end{figure}

Figure~\ref{fig:tel_thermal} illustrates the dynamics for the thermal
channel at $T=0.03\,\mathrm{K}$.  The left panel shows $F_A(\tau)$
alongside $C(\tau)$, $\mathcal{N}_1(\tau)/2$, and $\mathrm{ASC}(\tau)/2$
as functions of the dimensionless time $\tau=\kappa t$.  The fidelity
crosses the classical limit $2/3$ at the exact moment concurrence
vanishes ($\tau=\kappa t_{\rm ESD}$, dotted vertical line), confirming
Eq.~(\ref{eq:FA_C_equivalence}).  After this time, $\mathcal{N}_1$
remains frozen at $|b_3|$ and ASC decays toward the same asymptote,
yet $F_A$ falls below $2/3$.  The temperature dependence is shown
separately in Fig.~\ref{fig:tel_heatmap}: higher $T$ reduces both the
initial fidelity and the length of the teleportation window.

\begin{figure}[t]
    \centering
    \includegraphics[width=\columnwidth]{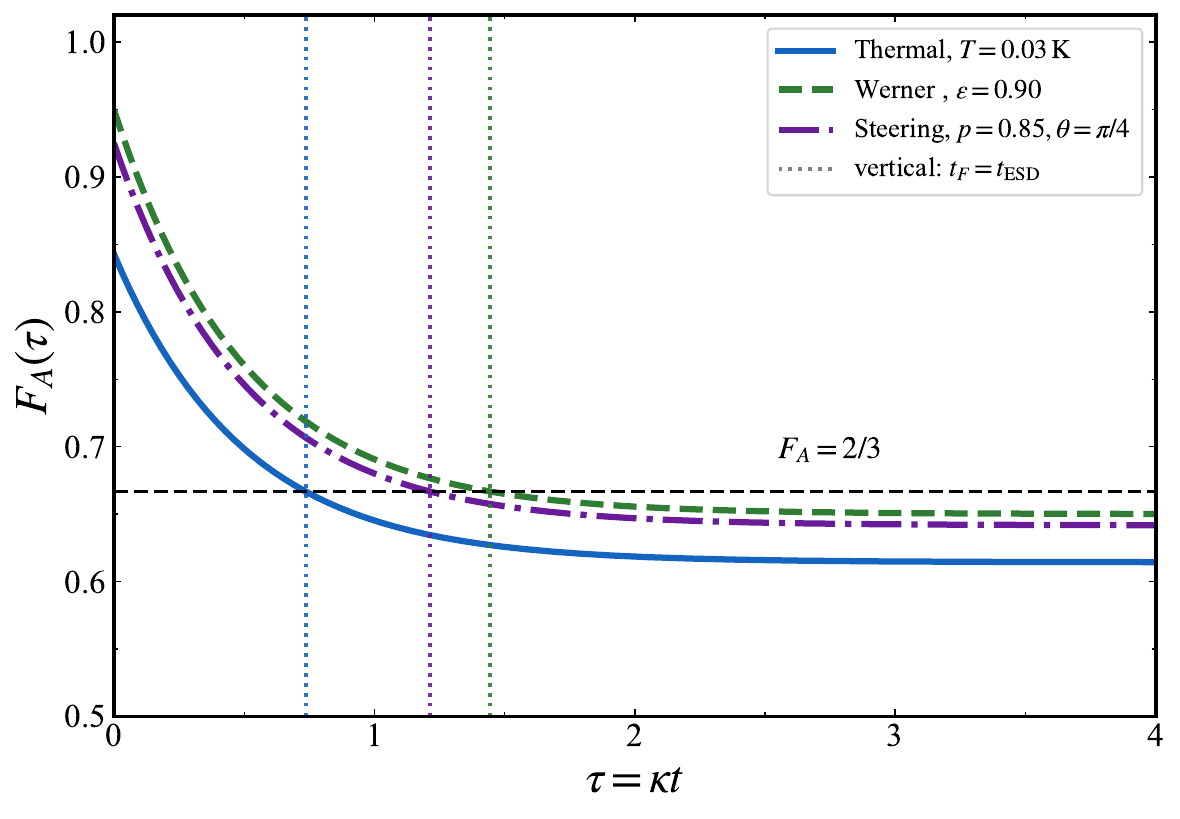}
    \caption{
Average teleportation fidelity for three channels: the thermal
hyperfine state ($T=0.03$~K), the Werner state $\varrho_-$
($\varepsilon=0.90$), and the one-way steering state at $\theta=\pi/4$
($p=0.85$). The dashed horizontal line is the classical bound
$F_A=2/3$. Each dotted vertical marks the teleportation lifetime
$\tau_F$, the rescaled time at which $F_A=2/3$, which by
Eq.~\eqref{eq:FA_C_general} is exactly where that state's concurrence
vanishes.} Every channel reaches the classical limit at finite time.

    \label{fig:tel_allstates}
\end{figure}

Figure~\ref{fig:tel_allstates} compares $F_A(\tau)$ (solid) and
$C(\tau)$ (dashed) for all three initial states: the thermal channel
at $T=0.03\,\mathrm{K}$, the Werner state $\varrho_-$ with
$\varepsilon=0.9$, and the one-way steering state with $p=0.85$,
$\theta=\pi/4$(kept at $\theta=\pi/4$ for the teleportation comparison specifically, for the reason given above; the correlation dynamics of Sec.~\ref{sec:results_steering} are shown separately at $\theta=0.4$, using the general derivation of Sec.~\ref{sec:oneway_deriv}).  In every case the fidelity reaches $2/3$ precisely
where the corresponding concurrence vanishes (dotted vertical markers),
confirming Eq.~(\ref{eq:FA_C_general}).

The joint dependence of the thermal channel fidelity on temperature
and dephasing is shown in Fig.~\ref{fig:tel_heatmap}.  The white
contour marks the classical threshold $F_A=2/3$, coinciding with
the ESD boundary $t_{\rm ESD}(T)$ from Eq.~(\ref{eq:tF_thermal}).
At low temperatures the system approaches the pure singlet and the
quantum-useful region extends to large $\tau$; at high temperatures
the thermal mixing reduces the initial coherence and the useful window
collapses toward $\tau=0$.

\section{Experimental Relevance: Pauli Tomography Protocol}
\label{sec:tomography}

All three correlation measures can be reconstructed from three joint Pauli
expectation values, without requiring full quantum state tomography. A general
two-qubit density matrix has 15 independent real parameters; for the Bell-diagonal X-state
family these reduce to 3 (the correlation coefficients $b_1,b_2,b_3$; the four populations are fixed by $b_3$ alone, Eq.~\eqref{eq:Xstate}), and all three correlation measures depend only on the three
two-body Pauli correlators defined below. This reconstruction is valid only within the Bell-diagonal family: a general X-state has additional independent parameters -- at minimum the local magnetizations $\langle\sigma_z\otimes\mathbb I\rangle$ and $\langle\mathbb I\otimes\sigma_z\rangle$, both identically zero only for maximally-mixed-marginal states -- which are not determined by $\langle XX\rangle,\langle YY\rangle,\langle ZZ\rangle$ alone. For the one-way steering state, for instance, $\langle ZZ\rangle=p$ is still cleanly measurable and constant for any $\theta$ (Eq.~\eqref{eq:sw_ZZ}), but reconstructing $C$, $\mathcal N_1$, and ASC in full would additionally require the two time-dependent local magnetizations of Eq.~\eqref{eq:sw_local_bloch}.

We introduce the shorthand $\langle XX\rangle \equiv \langle\sigma_x\otimes\sigma_x\rangle$,
$\langle YY\rangle \equiv \langle\sigma_y\otimes\sigma_y\rangle$, and
$\langle ZZ\rangle \equiv \langle\sigma_z\otimes\sigma_z\rangle$ for the joint
two-spin Pauli expectation values. For the time-evolved X-state, these follow
directly from $c_i(t)$ in Eqs.~(\ref{eq:c1})--(\ref{eq:c3}) and the
solutions:
\begin{align}
    \langle ZZ\rangle &= b_3 \quad(\text{constant}),
    \label{eq:ZZ}\\
    \langle XX\rangle &= b_1\,e^{-2\kappa t},
    \label{eq:XX}\\
    \langle YY\rangle &= b_2\,e^{-2\kappa t}.
    \label{eq:YY}
\end{align}
Two consistency checks follow: $\langle ZZ\rangle$ is constant (population
conservation under pure dephasing), and the ratio $\langle XX\rangle/\langle YY\rangle
= b_1/b_2$ is time-independent under the symmetric dephasing model assumed here, so it serves as a consistency check on that model.

Substituting into the closed-form expressions:
\begin{align}
    C_-(t) &= 2\max\!\left\{0,\;
    \frac{|\langle XX\rangle - \langle YY\rangle|
          - (1-\langle ZZ\rangle)}{4}
    \right\},
    \label{eq:C_tomo_minus}\\[4pt]
    C_+(t) &= 2\max\!\left\{0,\;
    \frac{|\langle XX\rangle + \langle YY\rangle|
          - (1+\langle ZZ\rangle)}{4}
    \right\},
    \label{eq:C_tomo_plus}
\end{align}
with $C(t) = \max\{C_-(t),\,C_+(t)\}$; and
\begin{align}
    \mathcal{N}_1(t) &= \max\!\left\{|\langle XX\rangle|,\,
    |\langle YY\rangle|,\,|\langle ZZ\rangle|\right\},
    \label{eq:Tmin_tomo}\\[4pt]
    \mathrm{ASC}(t) &=
    |\langle XX\rangle| + |\langle YY\rangle| + |\langle ZZ\rangle|.
    \label{eq:ASC_tomo}
\end{align}
The saturation transition of Trace MIN is directly readable from
Eq.~(\ref{eq:Tmin_tomo}): once $|\langle XX\rangle|$ and $|\langle YY\rangle|$
both fall below $|\langle ZZ\rangle|=|b_3|$, the maximum is set entirely by
the constant $\langle ZZ\rangle$.

\begin{figure*}[tp]
  \centering \includegraphics[width=\textwidth]{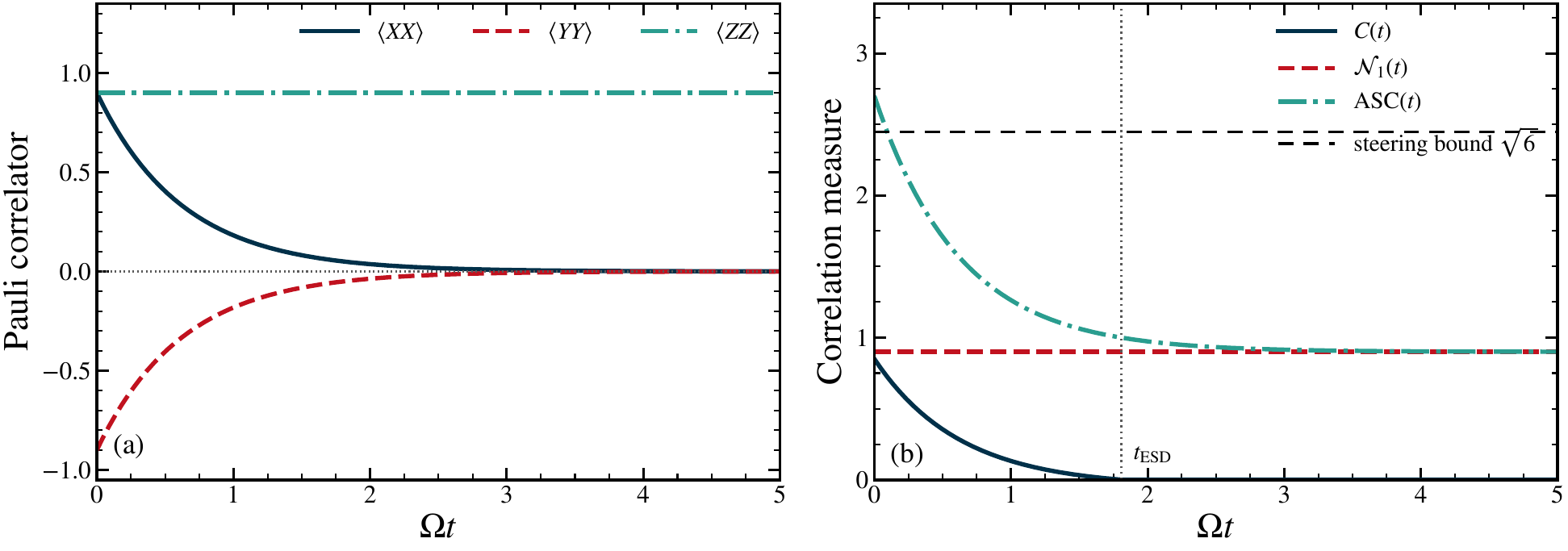}
    \caption{Pauli tomography for $\varrho_+$ at $\varepsilon=0.9$,
    $\kappa=0.8\,\alpha$. \textit{Left}: the three measured correlators,
    with $\langle ZZ\rangle=0.9$ constant. \textit{Right}: $C$,
    $\mathcal{N}_1$ and $\mathrm{ASC}$ rebuilt from them alone via
    Eqs.~(\ref{eq:C_tomo_minus})--(\ref{eq:ASC_tomo}). This panel plots
    $\mathrm{ASC}$ itself, so the dashed line sits at
    $\sqrt6\approx2.449$; the vertical marks $\Omega t_{\rm ESD}=1.81$.}
    \label{fig:tomo_panel}
\end{figure*}

Figure~\ref{fig:tomo_panel} verifies the protocol for $\varrho_+$ with
$\varepsilon = 0.9$ and $\kappa = 0.8\,\alpha$. The constant
$\langle ZZ\rangle = 0.9$ directly gives the Trace MIN asymptote.

\subsection{Implementation in Atomic Hydrogen}

\paragraph{Step 1 --- State preparation.}
Prepare the electron--proton spin pair using optical pumping via Lyman-$\alpha$
radiation (121.6\,nm), followed by resonant RF pulses on the 1420\,MHz
hyperfine transition to set the desired $b_1, b_2, b_3$~\cite{Maleki2021}.

\paragraph{Step 2 --- Controlled dephasing.}
Realize the dephasing channel via engineered longitudinal ($z$-axis) magnetic field noise
of controlled spectral density~\cite{Sheludiakov2019}, as modeled by the pure-dephasing Lindblad operators $F_e=\sigma_z\otimes\mathbb I$, $F_p=\mathbb I\otimes\sigma_z$ of Eq.~\eqref{eq:dissipator}. This is a phenomenological modeling choice, not a first-principles derivation from a specified noise-field geometry; transverse ($x/y$-axis) field fluctuations would instead induce transitions ($T_1$-type relaxation) rather than pure dephasing.

\paragraph{Step 3 --- Pauli spin measurements.}
Measure $\langle ZZ\rangle$, $\langle XX\rangle$, $\langle YY\rangle$ by
spin-resolved detection (Lyman-$\alpha$ fluorescence for the electron spin,
NMR for the nuclear spin), rotating transverse correlators into the $\hat{z}$
basis via $\pi/2$ pulses.

\paragraph{Step 4 --- Reconstruction.}
Compute $C$, $\mathcal{N}_1$, and ASC from
Eqs.~(\ref{eq:C_tomo_minus})--(\ref{eq:ASC_tomo}) and check the two
consistency conditions.

In solid H$_2$ films at $T < 1$\,K with $N \sim 10^{10}$ atoms,
shot-noise sensitivity $\sim 1/\sqrt{N}\sim 3\times10^{-5}$ is well
below the $\sim 10^{-1}$-scale signals here~\cite{Sheludiakov2019}.

\section{Conclusions}
\label{sec:conclusions}

We have investigated the dynamics of concurrence, Trace~MIN, and average steering coherence in the hydrogen hyperfine spin system under local dephasing by solving the Lindblad master equation exactly for the X-state family. Across all initial states considered, the three quantities preserve the hierarchy
$
C(t)\leq \mathcal{N}_1(t)\leq \mathrm{ASC}(t),
$
showing a clear separation in the robustness of different forms of quantum correlations.

For the Bell-diagonal X-state family, the long-time behavior is governed by the dephasing-invariant correlator $b_3=\langle\sigma_z\otimes\sigma_z\rangle$: Trace~MIN and ASC saturate at $|b_3|$ rather than vanishing, while concurrence disappears at a finite time through entanglement sudden death (the hyperfine singlet being the limiting case that decays asymptotically). The independence of the frozen Trace~MIN value from the dephasing rate is consistent with earlier studies of discord freezing~\cite{Mazzola2010,Cianciaruso2015,Bromley2015}, and our results indicate that a similar behavior extends to steering-based quantum correlations in this physical setting.

For the one-way steering state, whose marginals are not maximally mixed for $\theta\ne\pi/4$, we solved the full Lindblad equation with the Hamiltonian term retained and found a related but distinct mechanism: the population correlator $\langle ZZ\rangle=p$ is exactly constant for any $\theta$, while the population imbalance between $|01\rangle$ and $|10\rangle$ executes a damped oscillation at frequency $\approx4\Omega$, decaying at rate $\kappa$ rather than $2\kappa$. This is a genuinely hyperfine-driven feature, absent from the Bell-diagonal family, where $\Omega$ does not enter the dynamics at all. Because this state's local Bloch vectors are nonzero, Trace~MIN and ASC do not admit the simplified closed forms used elsewhere in this work and were evaluated from their operational definitions. For $\theta\ne\pi/4$, Trace~MIN does not freeze despite the constancy of $\langle ZZ\rangle=p$. Freezing therefore requires the additional condition of maximally mixed marginals.

We additionally examined the ability of the dephased thermal hyperfine state to act as a quantum teleportation channel. Using the Horodecki criterion, we obtained an analytical expression for the average teleportation fidelity and showed that the teleportation advantage region is exactly identical to the entanglement survival region,
$
F_A> \frac{2}{3}
\Longleftrightarrow
C>0,
$
for the considered family of X states with locally maximally mixed marginals (the thermal channel, the Werner states, and the one-way steering state at $\theta=\pi/4$). The persistence of Trace~MIN and ASC beyond entanglement sudden death does not translate into an advantage for the standard teleportation protocol considered here.

Finally, we proposed an experimentally accessible reconstruction scheme based on Pauli spin correlators that avoids full state tomography and requires only a small set of joint measurements.

The closed-form results derived here cover the Bell-diagonal (maximally-mixed-marginal) X-state family, Eq.~\eqref{eq:Xstate}, together with the more general one-way steering state at arbitrary $\theta$, for which $C(t)$ is closed-form and $\mathcal N_1(t)$, $\mathrm{ASC}(t)$ are evaluated numerically (Sec.~\ref{sec:oneway_deriv}). The broader X-state family studied by Mishra, Thapliyal, Pathak, and Venugopalan~\cite{Mishra2019} and Mishra, Thapliyal, and Pathak~\cite{Mishra2022}, including maximally nonlocal mixed states, also provides a useful setting to examine the hierarchy and freezing behavior beyond the two cases considered here.

\end{document}